\documentclass[aps,prb,twocolumn,floats,epsfig]{revtex4}
\usepackage{amssymb}
\usepackage{amsbsy}
\usepackage{amsmath}
\usepackage{epsfig}
\usepackage{color}
\usepackage{float}
\usepackage[colorlinks,linktocpage,bookmarks=false,citecolor=blue,linkcolor=red,urlcolor=blue]{hyperref}

\newcommand{\bib}{\bibitem}
\newcommand{\beq}{\begin{equation}}
\newcommand{\eeq}{\end{equation}}
\newcommand{\bea}{\begin{eqnarray}}
\newcommand{\eea}{\end{eqnarray}}

\newcommand\eq[1]{Eq.\ \ref{#1}}

\begin{document}

\title{Conformal Floquet dynamics with a continuous drive protocol}

\author{Diptarka Das$^{(1)}$, Roopayan Ghosh$^{(2)}$, and K. Sengupta$^{(2)}$}
\email{didas@iitk.ac.in, tprg@iacs.res.in, tpks@iacs.res.in }
\affiliation{$^{(1)}$Department of Physics, Indian Institute of
Technology
Kanpur, Kanpur 208016, India \\
$^{(2)}$School of Physical Sciences, Indian Association for the
Cultivation of Science, 2A and 2B Raja S. C. Mullick Road, Jadavpur,
Kolkata 700032, India}

\date{\today}

\begin{abstract}

We study the properties {of a conformal field theory}
(CFT) driven periodically with a continuous protocol characterized
by a frequency $\omega_D$. Such a drive, in contrast to its discrete
counterparts (such as square pulses or periodic kicks), does not
admit exact analytical solution for the evolution operator $U$. In
this work, we develop a Floquet perturbation theory which provides
an analytic, albeit perturbative, result for $U$ that matches exact
numerics in the large drive amplitude limit. We find that the drive yields the
well-known heating (hyperbolic) and non-heating (elliptic) phases
separated by transition lines (parabolic phase boundary). Using this
and starting from a primary state of the CFT, we compute the return
probability ($P_n$), equal ($C_n$) and unequal ($G_n$) time
two-point primary correlators, energy density($E_n$), and
the $m^{\rm th}$ Renyi entropy ($S_n^m$) after $n$ drive cycles. Our
results show that below a crossover stroboscopic time scale $n_c$,
$P_n$, $E_n$ and $G_n$ exhibits universal power law behavior as the
transition is approached either from the heating or the non-heating
phase; this crossover scale diverges at the transition. We also
study the emergent spatial structure of $C_n$, $G_n$ and $E_n$ for
the continuous protocol and find emergence of spatial divergences of
$C_n$ and $G_n$ in both the heating and non-heating phases. We express
our results for $S_n^m$ and $C_n$ in terms of conformal blocks and
provide analytic expressions for these quantities in several
limiting cases. Finally we relate our results to those obtained from
exact numerics of a driven lattice model.

\end{abstract}

\maketitle

\section{Introduction}
\label{intro}

The study of driven quantum systems have attracted a lot of
attention in recent years \cite{rev1,rev2,rev3,rev4}. The
theoretical interest in this area stemmed from the fact that such
systems provide access to a gamut of phenomena that have no analog
in their equilibrium counterparts. Some of these phenomena, for
periodically driven systems, include dynamical phase transitions
\cite{heyl1, sen1}, dynamical freezing \cite{das1, pekker1, sen2},
realization of time crystals \cite{ach1,av1}, and the possibility of
tuning ergodicity of the driven system with the drive frequency
\cite{sen3}. Moreover, periodically driven systems can lead to novel
steady states which has no counterpart in equilibrium quantum
systems \cite{ach2, sen2}. The study of these phenomena has also
received significant impetus from the possibility of realization of
closed quantum systems using ultracold atom platforms; indeed, such
platforms have recently been used to experimentally probe several
non-equilibrium phenomena \cite{rev5, exp1,exp2}.

A large set properties of such periodically driven systems can be
inferred from studying its Floquet Hamiltonin $H_F$ which is related
to the evolution operator $U$ of the system via the relation
$U(T,0)= \exp[-i H_F T/\hbar]$ \cite{rev3}. The computation of $H_F$
for a generic many-body system provides a significant challenge;
indeed, an exact analytic computation of $H_F$ is generally possible
for integrable models driven by discrete stepwise protocols (such a
square pulse or kicks) \cite{rev3}. This has led to development of
several perturbative schemes for computation of $H_F$; some of these
include Magnus expansion \cite{rev3}, adiabatic-impulse
approximation \cite{rev6}, the Hamilton flow method \cite{hamref},
and the Floquet perturbation theory (FPT) \cite{ds1,tb1}. Out of
these methods, the Floquet perturbation theory has the advantage of
being easily applicable to a wide class of systems as well as being
accurate over a large range of drive frequencies \cite{rg1}.

More recently, several theoretical studies concentrated on the
effect of Floquet dynamics on conformal field theories \cite{cft1,
cft2,cft3,cft4,cft5,cft6}. Usually driving a CFT is expected to
generate an additional scale in the problem which drives the system
away from its conformally invariant fixed point. However, it was
recently realized that there is a class of models \cite{ssdpapers,
vidalpaper} where drive protocols need not break the conformal
symmetry \cite{cft1}. It is found that for CFTs with a sine square
deformation (SSD) such dynamics can be initiated by a Hamiltonian
whose holomorphic part is given by
\begin{equation}
H(t) = \frac{2\pi}{L} \left[f(t) L_0
 + \frac{1}{2}f_1(t) (L_1+L_{-1}) \right],
\label{hamcft}
\end{equation}
where $L_n$ for each integer $n$ denotes a holomorphic generator
of the Virasoro algebra,
\begin{equation}
[L_m,L_n] = (m-n)L_{m-n} + \frac{c_0}{12} m(m^2-1)\delta_{m+n,0},
\label{vira}
\end{equation}
where $c_0$ denotes the central charge. These  holomorphic
generators are related to the stress tensor $T_{\mu\nu}$ of the CFT
by
\begin{equation}
\label{virstr}
\begin{split}
L_0 &= \frac{L}{2\pi} \int_0^L  dx  T_{00}(w) \\
L_{\pm 1} &= \frac{L}{2\pi} \int_0^L dx e^{\pm 2 \pi w /L} T_{00}(w)
\end{split}
\end{equation}
where $w=\tau +i x$, $x$ is the spatial coordinate and $\tau$ is the
Euclidean time. The anti-holomorphic ones have a similar expression
in terms of the complex conjugates.

It is well-known that the class of Hamiltonians given by Eq.\
\ref{hamcft} are valued in $su(1,1)$, giving rise to the evolution
operator \cite{cft1}
\begin{equation}
U(T,0) = {\mathcal T}\ e^{-\tfrac{i}{\hbar} \int\limits_0^T H(t)
dt}= \begin{pmatrix}a & b \\ c & d
\end{pmatrix}, \label{uexp1}
\end{equation}
valued in the group $SU(1,1)$ with $ad-bc=1$. Using the isomorphism
of $SU(1,1)$ and $SL(2,\mathbf{R})$ we shall also have occasion to
write the evolution operator in imaginary time in the generic form
\begin{equation}
U(T=-i \tau,0) = \begin{pmatrix}\tilde a & \tilde b\\ \tilde c &
\tilde d
\end{pmatrix}\in SL(2,\mathbf{R}).
\end{equation}
Its action on the complex plane $\mathbf{C}$ is given by the
M\"obius transformation
\begin{equation}
\label{moeb} z\to z'=\frac{\tilde a z+\tilde b}{\tilde c z+\tilde
d}, \quad z\in\mathbf{C}
\end{equation}
where $\tilde a,\tilde b,\tilde c,\tilde d \in\mathbf{R}$ and
$\tilde a \tilde d- \tilde b \tilde c$=1.

The exact solution for $U(T,0)$ for Hamiltonians valued in $su(1,1)$
and driven by discrete protocols were discussed earlier for periodic
kicks \cite{prapaper1} and square pulse protocols
\cite{cft1,cft2,cft3,cft4,cft5,cft6}. It was found that such driven
system could display two distinct phases depending on the drive
parameters; these are termed as the heating (hyperbolic) and the
non-heating (elliptic) phase and are found to be separated by a
transition line (parabolic phase boundary) \cite{prapaper1}. The
presence of these phases can be shown to be a direct consequence of
non-compact nature of the SU(1,1) group. However, such studies have
not been extended to continuous drive protocols where exact analytic
results are not available \cite{comment1}; in particular, the phase
diagram of such a driven system for continuous drive protocols has
not been studied so far.

It was noted in Ref.\ \onlinecite{cft1} that the action of the
evolution operator $U$ on a primary operator of the CFT can be
understood, in the Heisenberg picture, in terms of a M\"obius
transformation of it's coordinates leading to
\begin{equation}
U^{\dagger}(T,0) O(z,{\bar z}) U(T,0) = \left(\frac{\partial
z'}{\partial z}\right)^{h} \left(\frac{\partial {\bar z'}}{\partial
{\bar z}}\right)^{{\bar h}} O(z',{\bar z'}), \label{opevol1}
\end{equation}
where $z'$ is defined in \eq{moeb} and the bar designates
anti-holomorphic variables throughout. Using this relation, and a
straightforward mapping from cylindrical or strip geometries to the
complex plane following standard prescription \cite{book1}, several
results were obtained on the energy density, correlation function
and entanglement entropy of the driven system
\cite{cft1,cft2,cft3,cft4,cft5,cft6}. It was shown that in the
heating phase, such a drive leads to emergent spatial structure of
the energy density \cite{cft5}. Moreover, it was found that the time
evolution of the entanglement entropy shows linear growth with $n$
(in the large $n$ or long-time limit) in the heating phase. In
contrast, it shows an oscillatory behavior in the non-heating phase
and a logarithmic growth on the parabolic phase boundary. All of
these studies focussed on evolution starting from the CFT vacuum on
a strip geometry \cite{cft1,cft2,cft3,cft4,cft5,cft6}; the dynamics
of the system starting from asymptotic states corresponding to
primary operators of the theory, {which necessitates
computation of four-point correlation functions of the primary
fields of the driven CFT}, has not been studied so far.

In this work, we study dynamics of conformal field theories
subjected to a continuous drive protocol. The protocol we use
corresponds to the Hamiltonian given by Eq.\ \ref{hamcft} with
$f_1(t)=1$ and
\begin{eqnarray}
f(t) = f_0 \cos(\omega_D t) + \delta f \label{protocol1}
\end{eqnarray}
where $f_0$ is the drive amplitude and $\delta f$ is a constant
parameter. In this study we shall focus on the large drive amplitude
regime which corresponds to $f_0 \gg \delta f, 1$. In this regime
for $\omega_D \ge \delta f, 1$, the FPT is expected to be accurate
and we expect this to provide us with an analytic, albeit
perturbative, understanding of the properties of the driven system.

The central results that we obtain from such a study are as follows.
First, we chart out the phase diagram of the driven system as a
function of $\delta f$ and $\omega_D$ using exact numerics. Our
results show re-entrant heating and non-heating phases separated by
a parabolic phase boundaries as a function of $\delta f$ and $T$.
Second, we provide analytic expressions of the evolution operator
$U(T,0)$ using FPT. The phase diagram obtained from the perturbative
result provides a near-exact match with its exact numerical
counterpart over a wide range of frequencies. Our perturbative
results indicate the existence of a parameter $\alpha$, given within
first order FPT by (where we have put $\hbar=1$)
\begin{eqnarray}
\alpha &=& \sum_{n=-\infty}^{\infty}  J_n \left( \frac{2f_0 \pi}{ L
\omega_D}\right) \frac{T}{(n \pi + \pi \delta f T/L)},
\label{alphaeq}
\end{eqnarray}
which indicates proximity of the system to the phase boundary. The
system stays in the non-heating phase for $\alpha^2 <1$ and heating
phase for $\alpha^2>1$; the phase boundary between these two phases
is given by $\alpha=\pm 1$. We also use the first order FPT results
to explain the re-entrant transitions between the heating and the
non-heating phases in the phase diagram. Third, we use the obtained
expressions for $U$ within FPT, to obtain analytic expressions for
energy density, equal and unequal time two-point correlation
functions, entanglement entropy, and return probability of generic
sine-square deformed CFT starting from a primary state. Our results
expresses these quantities in terms of $\alpha$ and allows us to
characterize their behavior near the transition from the heating
phase. For example, we find that the return probability of any
primary state after $n$ drive cycles shows an universal behavior
below a crossover time $n_c$ both in the heating and {the non-heating phases} 
near the transition line; for $n>n_{c}$, the
probability decays exponentially for the heating phase and remains
an oscillatory function in the non-heating phase. We provide
analytic estimate of $n_{c}$ as a function of $\alpha$. Fourth, we
discuss the nature of the emergent spatial structure of the energy
density and the correlation function of primary operators for the
drive protocol. We analytically show that the energy density of any
primary state in the heating phase displays peaks which shifts from
$L/4$ and $3L/4$ to $L/2$ as one moves from deep inside the heating
phase to the transition line. For unequal-time correlation function
starting from the CFT vacuum, we find a line of such peaks both in
the heating and the non-heating phases; the position of these peaks
can be analytically found within first order FPT. Finally, we
provide expressions of the equal-time correlation function $C_n$ and
half-chain $m^{\rm th}$ Renyi entropy $S_n^m$ of the driven CFT
after $n$ drive cycles starting from an initial primary state in
terms of conformal blocks ${\mathcal V}_p$. In general, these blocks
do not have analytical expression for arbitrary CFTs; here, we
provide their analytical forms in several asymptotic limits. We
discuss the applicability of these limits to the driven CFT and
discuss the properties of $C_n$ and $S_n^m$ in these asymptotic
regimes. Finally, we relate some of our results to those obtained by
exact numerical study of the SSD model on a 1D lattice
\cite{ssdpapers, vidalpaper}.

The plan of the rest of the paper is as follows. In Sec.\
\ref{secdrive}, we derive expressions of $U$ and provide the phase
diagram for our drive protocol. This is followed by Sec.\
\ref{seccft} where we compute energy density, correlation functions
and return probabilities starting from a primary state. Next, in
Sec.\ \ref{secrel}, we relate our results to those obtained from
numerical study of driven SSD Hamiltonian on a 1D lattice. Finally,
we discuss our main results and conclude in Sec.\ \ref{conc}. Some
details of the perturbative FPT calculations and representation
independent derivation of the Mobius transformation are presented in
the Appendices.

\section{Phase Diagram}
\label{secdrive}

To find $U(T,0)$ corresponding to the Hamiltonian given by Eq.\
\ref{hamcft} with $f(t)$ given by Eq.\ \ref{protocol1}, we first
note that the holomorphic generators (Eq.\ \ref{vira}) for
$n=-1,0,1$ form an $su(1,1)$ subalgebra
\begin{equation}
\label{vir1} [L_0,L_{-1}] = L_{-1},\quad [L_0,L_1] = -L_1,\quad
[L_{-1},L_1]=-2L_0.
\end{equation}
A representation of this algebra is furnished in terms of the Pauli
matrices as
\begin{equation}
\label{repn} L_0=\tfrac{1}{2}\sigma_z,\quad L_{\mp 1} =
\pm\sigma^{\mp}, \quad\sigma^{\pm}=\frac{1}{2}(\sigma_x\pm
i\sigma_y),
\end{equation}
where $\sigma_{\alpha}$ for $\alpha=x,y,z$ are standard Pauli
matrices.
In this representation (Eq.\ \ref{repn}), the holomorphic part of
the Hamiltonian $H(t)$ becomes
\begin{eqnarray}
H(t) &=& \frac{\pi}{L} \left( f(t) \sigma_z - i \sigma_y \right)
\label{hampauli}\
\end{eqnarray}
which corresponds to a Zeeman Hamiltonian of a single
spin-half particle with a time dependent magnetic field along $\hat
z$ and an imaginary constant magnetic field along $y$. The latter
feature is a consequence of the SU(1,1) group structure and is
crucial in realization of the heating and the non-heating phases
\cite{prapaper1}. The expression for  $U(T,0)$ corresponding to Eq.\
\ref{hampauli} can be found exactly for periodic kicks
\cite{prapaper1} and square-pulse \cite{cft1} protocols.

\begin{figure}
\rotatebox{0}{\includegraphics*[width= 0.48 \linewidth]{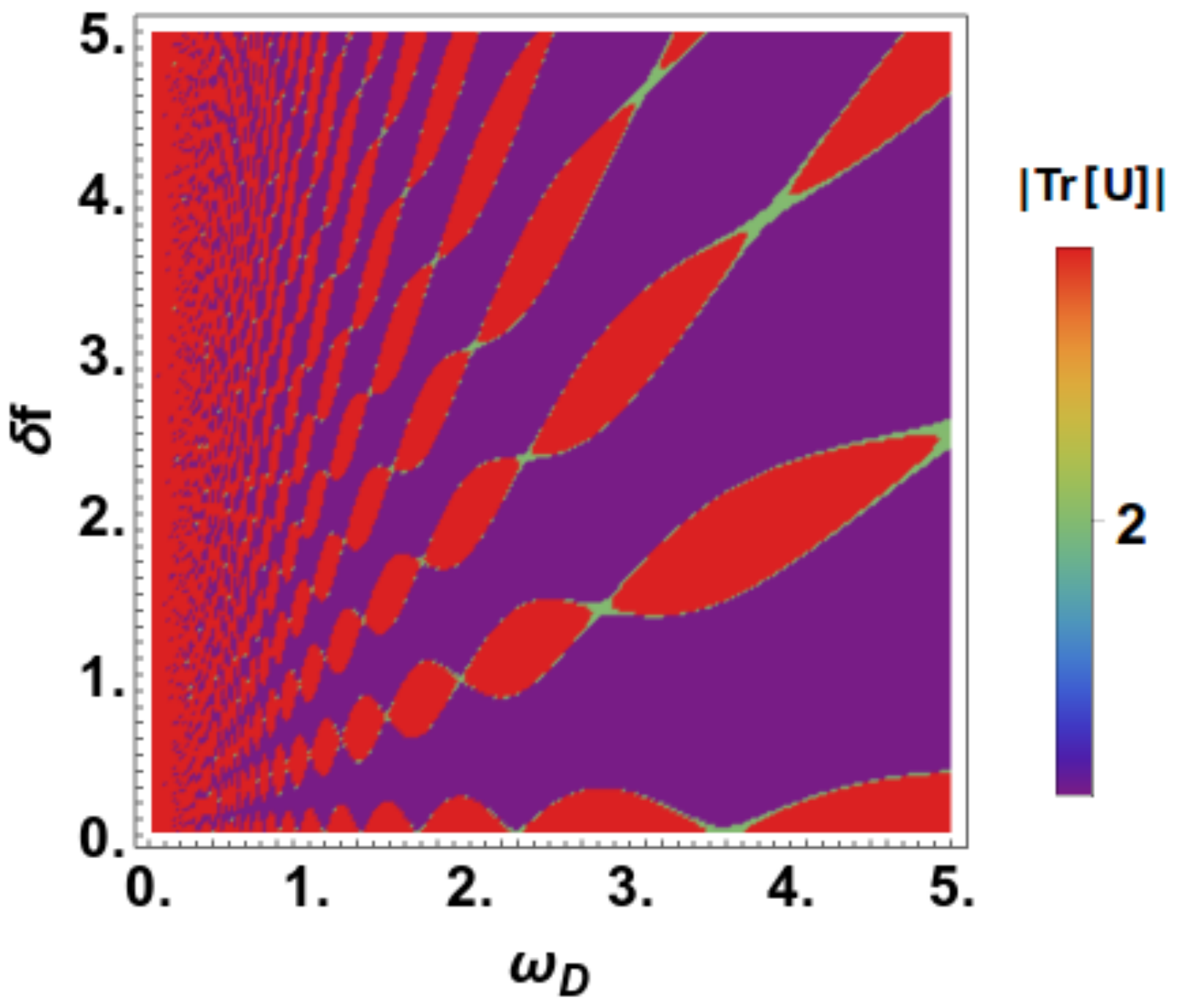}}
\rotatebox{0}{\includegraphics*[width= 0.48 \linewidth]{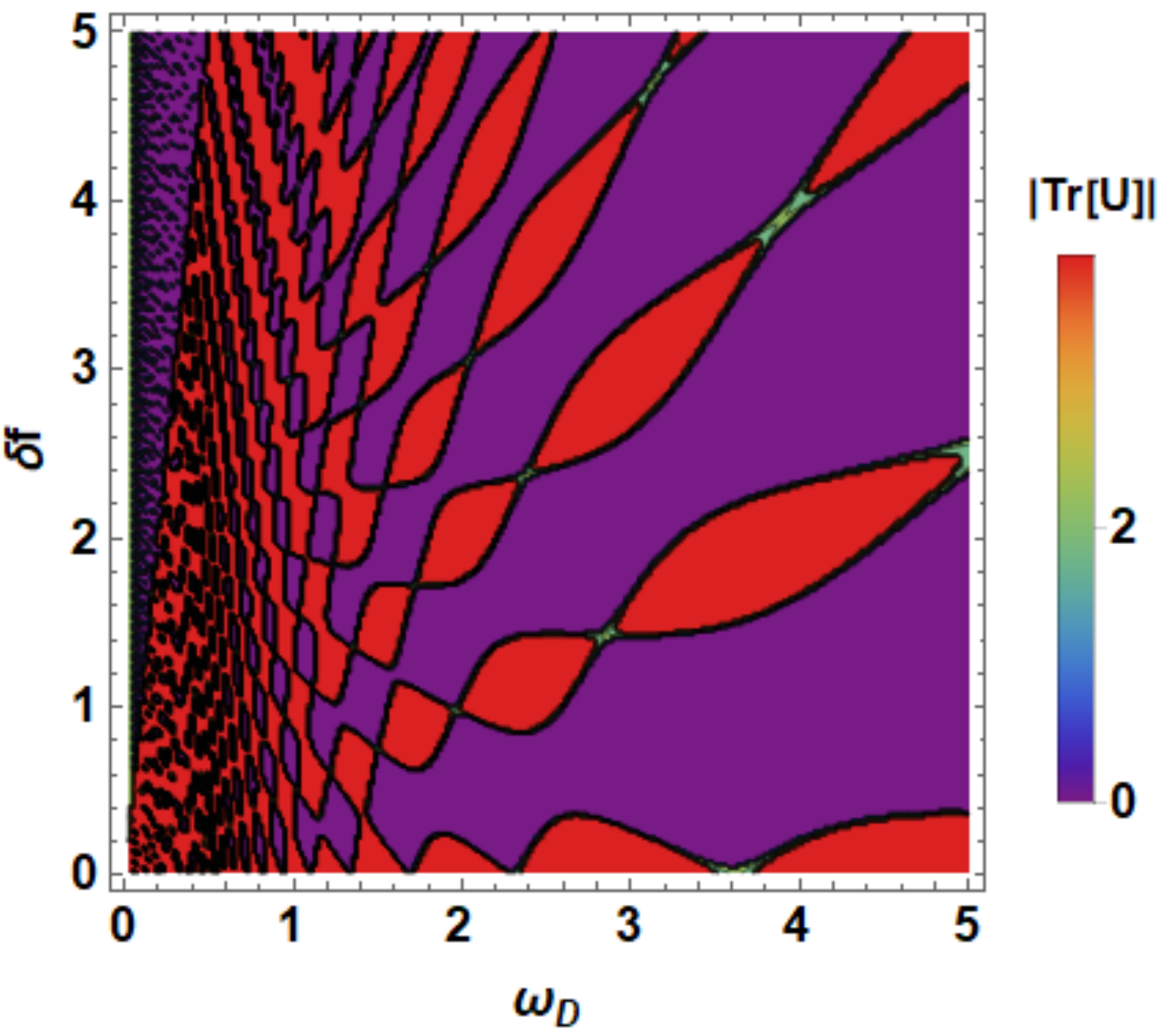}}
\caption{(Color online) Left Panel: Plot of the phase diagram,
obtained from $|{\rm Tr}U(T,0)|$ plotted as a function of the
amplitude $\delta f$ and frequency $\omega_D$ as obtained from exact
numerics. The red regions indicate heating phases while the violet
ones show the non-heating phases. The re-entrant transition between
these phases are shown by black lines indicating parabolic
transition lines. Right Panel: Similar phase diagram obtained from
first order FPT. For both plots, we have set $\pi/L$ to unity and
set $f_0=10$.} \label{fig1}
\end{figure}

In contrast, for the continuous protocol given by Eq.\
\ref{protocol1}, an exact analytic expression for $U(T,0)$ does not
exist. However, the numerical result for $U$  can be obtained in a
straightforward manner. Such a numerics is carried out by dividing
the time period $T$ into $N$ Trotter steps with width $\delta t_i
\sim T/N$. The maximal allowed width of each of these steps depend
on system energy scale and the drive frequency; they are chosen so
that $H$ does not vary appreciably within any step. This allows one
to write $U \simeq \prod_{i=1,N} U_i$, where $U_j= U(t_{j-1}+\delta
t_j,t_{j-1}) = \exp[-i \int_{t_{j-1}}^{t_{j-1}+\delta t_j} dt H(t)
/\hbar]$. This procedure leads to $U(T,0)= \exp[-i H_F T/\hbar]$; we
find numerically that the Floquet Hamiltonian is
\begin{eqnarray}
H_F &=& p(T) \sigma_z + i q(T)  \sigma_y \label{unum1}
\end{eqnarray}
We note that $\sigma_x$ does not appear in $H_F$. The condition for
the different phases are thus obtained by \cite{prapaper1} {$|{\rm Tr} U(T,0)| >2$ 
(heating phase), $<2$ (non-heating
phase), and $=2$ (phase boundary).} The exact numerical phase
diagram obtained from this procedure is shown in the left panel of
Fig.\ \ref{fig1}. We find several re-entrant transitions between the
heating and the non-heating phases leading to multiple lobes whose
boundary correspond to the transition lines. We note that the
transition between these phases can be induced by tuning either
$\delta f$ or $\omega_D$. In particular, we note that large $\hbar
\omega_D \gg \delta f ,1$, the phase transition between the heating
and non-heating phase occurs at $\delta f=1$.

The absence of $\sigma_x$ in $H_F$ can be shown to be the
consequence of an emergent dynamic symmetry of the evolution
operator at stroboscopic times which follows from the periodicity of
$H(t)$. To see this we note that since {$H(t)=
H(T-t)$}, one has $U_j= U_{N-j+1}$ for all $j$ in the Trotter
product. Further, we note for any of these Trotter steps, one has
$\sigma_x U_j \sigma_x = U_j^{-1}$. Thus one can write
\begin{eqnarray}
 \sigma_x U(T,0) \sigma_x &=& \sigma_x \left(\prod_{j=1,N} U_j \right) \sigma_x \label{syma} \\
&=& \prod_{j=1,N} U^{-1}_j = \left( \prod_{j=1,N} U_j \right)^{-1} =
U^{-1}(T,0) \nonumber
\end{eqnarray}
where we have used the relation $U_j^{-1} = U_{N-1+j}^{-1}$. Eq.\
\ref{syma} clearly implies that if $U= \exp[-i H_F T]$ then
$\sigma_x H_F \sigma_x = -H_F$. This forbids presence of terms $\sim
\sigma_x$ in $H_F$. It is to be noted that this symmetry is absent
for $t \ne nT$ where $n$ is an integer.

Next, we develop an perturbative analytic expression of $U$ in the
limit when $f_0$ is the largest scale in the problem. Here the
diagonal term in $H(t)$ is treated exactly and the effect of the
off-diagonal term is taken into account within standard
time-dependent perturbation theory \cite{tb1}. In this scheme, the
first term for $U$ and $H_F$ is given by
\begin{eqnarray}
U_0(t,0) &=& e^{-i (\pi/L) (f_0 \sin \omega_D t/\omega_D + \delta f
t)
\sigma_z} \nonumber\\
H_F^{(0)} &=& s \sigma_z /T  \label{zerothorder}
\end{eqnarray}
where $s= \arccos [\cos(\delta f T \pi/L)]$. We note that
$H_F^{(0)}$ retain the periodic structure of $U_0$.

To obtain the first order correction to $U(T,0)$, we use the
standard result \cite{tb1}
\begin{eqnarray}
U_1(T,0) = U_0(T,0) (I- U'_1(T,0)) \label{firstorderu}
\end{eqnarray}
where $I$ denotes the $2 \times 2$ identity matrix and $U'_1$
denotes the first-order correction to $U$ in the interaction picture
given by
\begin{eqnarray}
U'_1(T,0) &=& -i \int_0^T  dt U_0^{\dagger}(t,0) H_1 U_0(t,0)
\label{firstint1}
\end{eqnarray}
where $H_1= - \frac{\pi}{L} i \sigma_y$ is the perturbative term of
the Hamiltonian. A simple calculation detailed in App.\ \ref{appa}
yields
\begin{eqnarray}
U'_1(T,0) &=& i \alpha e^{is} \sin(s) \sigma_y \nonumber\\
U_1(T,0) &=& \left ( \begin{array} {cc} e^{- i s} & - i \alpha \sin s \\
i \alpha \sin s  & e^{is} \end{array} \right) \label{u1eq}
\end{eqnarray}
where $\alpha$ is given by Eq.\ \ref{alphaeq}. We note that
$U_1(T,0)$ is not unitary; this is a well-known issue with
perturbation theory for $U$. In what follows we unitarize $U_1(T,0)$
as follows. We note that this can be done by first writing $(I-
U'_1(T,0)) \simeq \exp[-i H'_F T]$, where the matrix $H'_F= (i/T)
U'_1$. The evolution operator is then given by $U_0(T,0) \exp[- i
H'_F T]$ which is unitary. However, in the present case, we find
that evolution operator obtained by this method retains terms $\sim
\sigma_x$. This is clearly inconsistent with the dynamic symmetry
discussed in Eq.\ \ref{syma}. Thus we chose to unitarize $U_1(T,0)$
directly since it does not have any term $\sigma_x$. These two
alternative routes to obtaining $H_F$ coincides with each other in
the large $\omega_D$ limit, where $s \to 0$.

The unitarization of $U_1(T,0)$ given by Eq.\ \ref{u1eq} can be
achieved by writing
\begin{eqnarray}
U_1(T,0) &=& \exp[-i \theta ( \sigma_z n_z + \sigma_y n_y)], \quad
\theta = s \sqrt{1-\alpha^2}  \nonumber\\
n_z &=& \frac{1}{\sqrt{1-\alpha^2}}, \quad n_y = \frac{i
\alpha}{\sqrt{1-\alpha^2}} \label{fpt1}
\end{eqnarray}
Comparing Eqs.\ \ref{unum1} and \ref{fpt1}, we find that the first
order FPT result provides approximate analytic expressions for
$p(T)$ and $q(T)$ given by
\begin{eqnarray}
q/p &\simeq & \alpha, \quad  p \simeq s/T  \label{comp1}
\end{eqnarray}
In what follows we shall use Eq.\ \ref{comp1} to compare between
exact numeric and approximate analytic results.

The evolution operator obtained in Eq.\ \ref{fpt1} indicates several
interesting features. First, we find that $n_y$ is imaginary; this
is a direct consequence of the SU(1,1) group structure. Second, we
note that ${\rm Tr}[U_1(T,0)]= 2 \cos(\theta)$; thus the condition
$|{\rm Tr}[U_1(T,0)]| > (<) 2$ for realization of heating
(non-heating) phases translates to $\theta$ being imaginary(real).
This happens when $\alpha^2 >(<)1$ which leads to our identification
of $\alpha$ as the parameter whose value determines the phase the
system. Third, the parabolic phase boundary between these two phases
is given by $|{\rm Tr}[U_1(T,0)]|=2$ which leads to {$\theta=0$}. This is 
realized for $\alpha=\pm 1$ on the boundary
between the heating and non-heating phases. From the expression of
$\alpha$ in Eq.\ \ref{alphaeq}, we find that for $\omega_D \gg f_0$,
the condition $\alpha=1$ is satisfied for $\delta f=1$. This can be
verified directly from the exact phase diagram in the left panel of
Fig.\ \ref{fig1}. Fourth, the re-entrance of the heating and
non-heating phases as a function of $\delta f$ shown in the right
panel of Fig.\ \ref{fig1} displaying the phase diagram obtained from
first order FPT, can be easily explained by periodic nature of
$|\cos(\theta)|$ since the phases must repeat for $\theta \to \theta
+\pi$. We note the two phase boundaries in each of the lobes of this
phase diagram correspond to $\alpha=1$ (lower branch of the lobes)
and $\alpha=-1$ (upper branch of the lobes). The center of these
lobes correspond to $\alpha \to \infty$ which occurs along the lines
$\delta f /\omega_D = n_0/2$ for $n_0 \in Z$. This can be directly
seen from Eq.\ \ref{alphaeq} where the term in the sum corresponding
to $n=n_0$ diverges in this limit. Finally, a comparison between the
left and the right panels of Fig.\ \ref{fig1} shows that the two
phase diagrams match qualitatively for {$\hbar
\omega_D/(\pi/L) >1$}. This allows us to justify the use of the
analytical method for a qualitative understanding of the dynamics
within first order FPT. A computation of the second-order results of
FPT is carried out in App.\ \ref{appa}; we find that it retains all
features of the first order theory and provides a near-identical
phase diagram. In App.\ \ref{appb}, we chart out a representation
independent derivation of the first order perturbation results in
the high frequency limit which matches the results obtained here
using the SU(1,1) representation of the Virasoro generators.

Before ending this section , we obtain the elements of $U^n=
U(nT,0)$ (where $ n\in Z$ is an integer). Using Eqs.\ \ref{uexp1}
and \ref{unum1}, we find the elements of $U^j$ for any integer $j$
obtained using exact numerics in the non-heating phase (where $p>q$)
to be
\begin{eqnarray}
a_j &=& \cos(\sqrt{p^2-q^2}j T) - i \frac{p}{\sqrt{p^2-q^2}}
\sin(\sqrt{p^2-q^2}j T) \nonumber\\
b_j &=& \frac{i q}{\sqrt{p^2-q^2}} \sin(\sqrt{p^2-q^2}j T)
\label{numexp2}
\end{eqnarray}
where $d_j=a_j^{\ast}$ and $c_j=b_j^{\ast}$. For the heating phase,
for which $p<q$, one obtains
\begin{eqnarray}
a'_j &=& \cosh(\sqrt{q^2-p^2}j T) - i \frac{p}{\sqrt{q^2-p^2}}
\sinh(\sqrt{q^2-p^2}j T) \nonumber\\
b'_j &=& \frac{i q}{\sqrt{q^2-p^2}} \sinh(\sqrt{q^2-p^2}j T)
\label{numexp3}
\end{eqnarray}
For the transition line where $p=q$, a careful evaluation of the
limit shows
\begin{eqnarray}
a"_j &=& 1- i p j T, \quad b"_j = -i p j T \label{numexp4}
\end{eqnarray}
The corresponding analytic, first order FPT, expressions of elements
of $U(jT, 0)$ in terms of $\alpha$ and $s$ can be directly read off
from these equations by using the correspondence expressed in Eq.\
\ref{comp1}. For example for the transition line, the first order
FPT result is $a"_j= 1- i j s$ and $b"_j= -i j s$. Also, the
corresponding elements $\tilde a_j$, $\tilde b_j$, $\tilde c_j$ and
$\tilde d_j$ of the evolution operator in Euclidean time can be
obtained for each of these phase from Eqs.\ \ref{numexp2},
\ref{numexp3}, and \ref{numexp4} via standard analytic continuation
$T \to -i \tau$. We shall use these expressions in the next section
computation of return probability, energy density, correlation
functions and the entanglement entropy.

\section{Results for driven CFT}
\label{seccft}

It was pointed out in Ref.\ \onlinecite{cft1} that the operation of
the evolution operator $U$ on any primary operator of the CFT can be
understood in terms of a M\"obius transformation of its coordinates
and is therefore given by Eq.\ \ref{opevol1}. This is further
discussed in App.\ \ref{appb}. In this section, we shall use this
result to study the properties of the driven CFT with a cylindrical
geometry which corresponds to periodic boundary condition of a
driven 1D chain of length $L$. In our setup, the coordinate of the
cylinder is given by $w= \tau +i x$, where $x$ is the spatial
coordinate and $\tau$ is the Euclidean time; we use $w= (L/(2\pi))
\ln z$ for mapping such a cylinder to the complex plane. Thus for
any primary operator ${\mathcal O}(w,\bar w)$ on the cylinder, we
can write \cite{book1,cft1}
\begin{eqnarray}
U^{ n \dagger} {\mathcal O}(w,\bar w) U^n &=& \left(\frac{\partial
w}{\partial z} \right)^{-h} \left(\frac{\partial z_n}{\partial z}
\right)^{h}\left(\frac{\partial {\bar w}}{\partial {\bar z}}
\right)^{-{\bar h}} \nonumber\\
&& \times \left(\frac{\partial {\bar z}_n}{\partial {\bar z}}
\right)^{{\bar h}} {\mathcal O}( z_n, {\bar z}_n) \label{opevol2}
\end{eqnarray}
where $z_n$ is the transformed coordinates given by Eq.\ \ref{moeb}
and $\tilde a_n$, $\tilde b_n$, $\tilde c_n$, and $\tilde d_n$ are
the elements of $U^n$ in Euclidean time which can be obtained from
analytic continuation of $T \to -i \tau$ in Eqs.\ \ref{numexp2},
\ref{numexp3} and \ref{numexp4}. In all computations, we shall work
in Euclidean time and carry out the analytic continuation to real
time using $\tau= iT$ at the end of the calculation.

Apart from the primary operators we shall also use transformation
properties of the stress tensor. The holomorphic part of the stress
tensor denoted by $T(w)$ transforms, under a general coordinate
transformation, as  $T(w) \to T'(w')= (\partial w'/\partial w)^{-2}
(T(w) + S_{w',w})$, where $S$ is Schwarzian given by
\begin{eqnarray}
S_{w', w} &=& \frac{c_0}{12} \left[ \left(\frac{\partial^3
w'}{\partial w^3} \right) \left(\frac{\partial w'}{\partial w}
\right)^{-1}
\right. \nonumber\\
&& \left. -\frac{3}{2} \left( \left(\frac{\partial^2 w'}{\partial
w^2} \right) \left(\frac{\partial w'}{\partial w} \right)^{-1}
\right)^2\right] \label{schdef}
\end{eqnarray}
Here we note that $S_{z_n,z}=0$ since $S$ vanishes for any M\"obius
transformation and $S_{z,w}= c_0/(24 z^2)$ for transformation from a
cylinder to the complex plane. The transformation for the
antiholomorphic part of the stress tensor can be obtained in the
similar manner by replacing $(w', w) \to ({\bar w}', \bar w)$.

In what follows, we shall use these general results to obtain the
stroboscopic time ($n$) dependence of the return probability, energy
density, equal and unequal time correlation functions, and the
entanglement entropy of the driven CFT. The initial state for this
purpose is chosen to be an asymptotic in-state of the CFT denoted by
\begin{eqnarray}
|h,\bar h\rangle &=&  \lim_{w, {\bar w} \to -\infty} \phi(w,\bar w)
|0 \rangle \label{hstate1}
\end{eqnarray}
where $|0\rangle$ denotes the CFT vacuum, $\phi$ is a primary field
with dimension $(h,\bar h)$ and we note $w, {\bar w} \to -\infty$
corresponds to $\tau \to -\infty$ which in turn implies $z, {\bar z}
\to 0$ on the complex plane. This defines for us a primary state.
The corresponding out-states are given by
\begin{eqnarray}
\langle h,\bar h | &=&  \langle 0| \lim_{w, {\bar w} \to \infty}
\phi(w,\bar w) \label{hstate2}
\end{eqnarray}
which corresponds to $\tau \to \infty$ and hence $z,\bar z \to
\infty$ on the complex plane.

The reason for choosing these primary states are as follows. First,
we note that since $L_0$ and $L_{\pm 1}$ annihilates the CFT vacuum,
the vacuum state does not evolve. This renders all equal time
correlation functions of primary operators, entanglement entropy and
energy density to be fixed at their equilibrium values under action
of $U$; only the unequal-time correlation function shows non-trivial
dynamics. This feature of driven CFT usually compels one to work
with the strip geometry \cite{cft1,cft4} or use different M\"obius
transformation \cite{cft5} where one can obtain non-trivial time
evolution starting from the ground state. Here we use an alternative
approach by using the cylindrical geometry but starting from the
primary states of the CFT ($|h,\bar h\rangle$) as initial states;
these are the simplest states of the model which have non-trivial
dynamics under action of $U$.

\subsection{Return probability and energy density}
\label{rpensec}

The return probability amplitude $A_n$ after $n$ cycles of the drive
is given by
\begin{eqnarray}
A_n  &=& \frac{\langle h, \bar h| U(n T,0) |h , \bar
h\rangle}{\langle h, \bar h|h, \bar h\rangle|} \label{rpeq1}
\end{eqnarray}
where the denominator corresponds to the normalization of the
asymptotic states. To this end, we first look at the contribution of
the holomorphic part of the return probability amplitude given by
\begin{eqnarray}
A_n^{\rm hol} &=& \frac{\lim_{w_1 \to 0, w_2 \to \infty} \langle 0|
\phi(w_2) U(nT,0) \phi (w_1) |0\rangle}{\lim_{w_1 \to 0, w_2 \to
\infty} \langle 0| \phi(w_2) \phi (w_1) |0\rangle} = \frac{\mathcal
N}{\mathcal D} \nonumber\\ \label{rpa1}
\end{eqnarray}
To compute $A_n^{\rm hol}$, we take $w_i = \tau_i + i x_i$ for
$i=1,2$ and consider the limits $\tau_1 \to -\infty$ and $\tau_2 \to
\infty$ at the end of the calculation. Using the fact that $\langle
0| U^{\dagger} = \langle 0|$, we can rewrite as ${\mathcal N}=
\langle 0| U^{n \dagger} \phi(w_2, \bar w_2) U^n \phi(w_1,{\bar
w}_1)|0\rangle$. We then use Eq.\ \ref{opevol2} to obtain
\begin{eqnarray}
{\mathcal N} &=& \lim_{z_2 \to \infty, z_1 \to 0}  \left (\frac{2
\pi}{L}\right)^{2h} \left( \frac{z_1 z_2 }{(\tilde c_n z_2 +\tilde
d_n)^2(z_{n 2}
-z_1)^2}\right)^h \nonumber\\
{\mathcal D} &=& \lim_{z_2 \to \infty, z_1 \to 0}
\left(\frac{2\pi}{L}\right)^{2h} \left(\frac{z_1 z_2}{(z_2-z_1)^2}
\right)^{h} \label{rpa2}
\end{eqnarray}
where $z_{n2} = (\tilde a_n z_2+\tilde b_n)/(\tilde c_n z_2 + \tilde
d_n)$ is the transformed coordinate, $\tilde a_n$, $\tilde b_n$,
$\tilde c_n$ and $\tilde d_n$ are obtained from Eqs.\ \ref{numexp2},
\ref{numexp3} ,\ref{numexp4} after analytic continuation $T=-i \tau$
and we have used $\partial w/\partial z= L/(2\pi z)$ and $\partial
z_{n2}/\partial z= (\tilde c_n z_2+\tilde d_n)^{-2}$. After a
careful evaluation of the limits, we finally obtain $A_n^{\rm hol} =
\tilde a_n^{-2h}$. A similar computation shows the contribution from
the anti-holomorphic part to be $A_n^{{\rm ant-hol}}= \tilde a_n^{-2
\bar h}$. Thus one obtains the return probability $P_n$, after
analytic continuation to real time, to be
\begin{eqnarray}
P_n &=& |A_n^{\rm hol} A_n^{{\rm ant-hol}}|^2 = |a_{n}|^{-4(h+ \bar
h)} \label{rpeq2}
\end{eqnarray}
The exact numerical value of $P_n$ can thus be obtained by using
Eqs.\ \ref{numexp2}, \ref{numexp3}, \ref{numexp4}. Here and for all
quantities in the rest of this section, we shall provide expressions
for the prediction of FPT; the numerical result in terms of $p$ and
$q$ can be directly read from these expressions by using Eq.\
\ref{comp1}({\it i.e.} with the mapping $\alpha \to p/q$ and $s \to
p T$). This procedure yields {
\begin{eqnarray}
P_n &=&  \left( \frac{\alpha^2(1+\cosh n
\theta)-2}{2(\alpha^2-1)}\right)^{-2(h+\bar h)}, \, {\rm heating}
\nonumber\\
&=& \left( \frac{2- \alpha^2(1+\cos n
\theta)}{2(1-\alpha^2)}\right)^{-2(h+\bar h)}, \, {\rm non-heating}
\nonumber\\
&=& (1+ n^2 s^2)^{-2(h+\bar h)}, \quad {\rm transition \, line}
\label{rpana1}
\end{eqnarray}
}
\begin{figure}
\rotatebox{0}{\includegraphics*[width= 0.48 \linewidth]{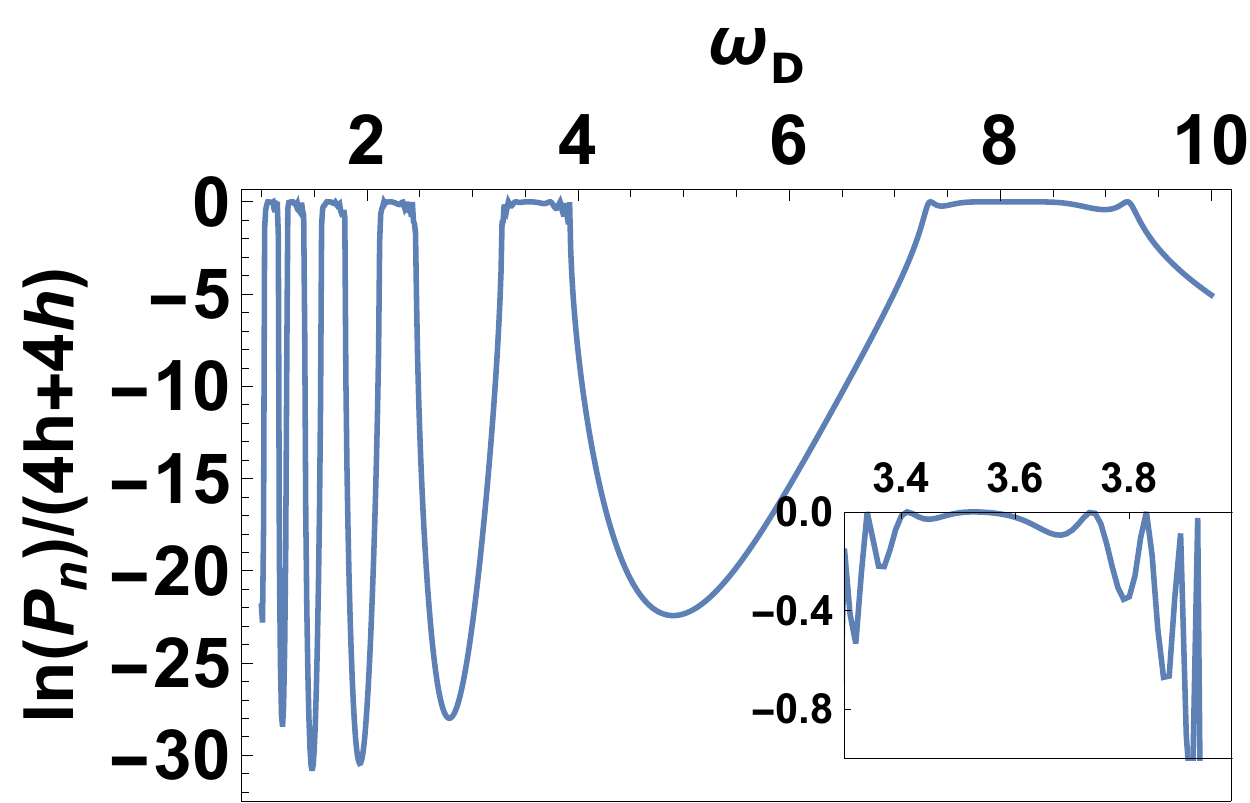}}
\rotatebox{0}{\includegraphics*[width= 0.48 \linewidth]{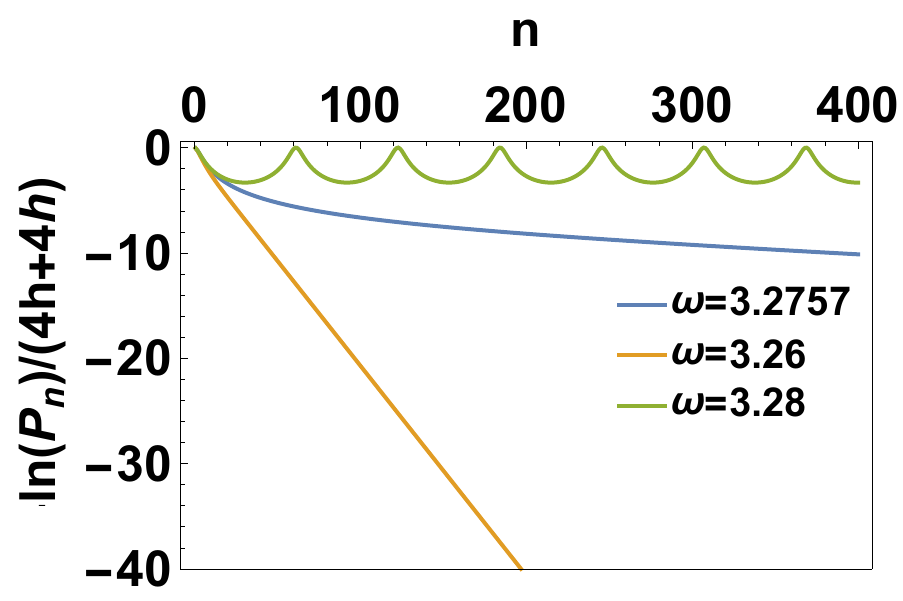}}
\caption{(Color online) Left Panel: Plot of $\ln P_n /(4(h+\bar h))$
as a function of frequency $\omega_D$ for after $n=50$ cycles of the
drive. The inset shows the variation of $\ln P_n$ with $\omega_D$ in
the non-heating phase. Right Panel: Plot of $\ln P_n /(4(h+\bar h))$
as a function of $n$ for three representative drive frequencies
corresponding to the heating ($\omega_D= 3.26$ orange line) and
non-heating phases ($\omega_D= 3.28$, green line) and the transition
line ($\omega_D \simeq 3.2767$, blue line). In all plots $\pi/L$ has
been set to unity, $f_0=10$ and $\delta f=0.5$. See text for
details.} \label{fig2}
\end{figure}
The behavior of $\ln P_n$ is shown in the left panel of Fig.\
\ref{fig2} as a function of drive frequency after $n=50$ cycles of
the drive and for $\delta f=0.5$. We find that $\ln P_n$ shows dips
at large $n$ in the heating phases; in contrast, it shows
oscillatory behavior near the transition in the non-heating phase as
can be seen from the inset of the left panel of Fig.\ \ref{fig2}. We
also note that the return probability shows an exponential decay for
large $n$ in the heating phase and an oscillatory behavior in the
non-heating phase according to standard expectation. On the
transition line, it shows a power law decay {$\sim
n^{-2(h+\bar h)}$} for large $n$. These qualitatively different
behaviors of $P_n$ can be clearly seen in the right panel of Fig.\
\ref{fig2} where $\ln P_n$ is plotted as a function of $n$ for three
values of drive frequencies corresponding to three different phases.
Here we also note that near the phase transition line, in either
phase, the behavior of $P_n$ is identical to that on the critical
line below a crossover timescale $n \le n_{c}$. This is most easily
seen by expanding the $ \cosh(n\theta)$ term in powers of $n \theta$
in Eq.\ \ref{rpana1}. This procedure yields, in the heating phase,
{\begin{eqnarray}
&& \ln P_n  \simeq  -2(h+\bar h)\ln \left[1+ (\alpha s n)^2 +
(\alpha s n)^4 \frac{\alpha^2-1}{3
\alpha^2} + ... \right] \nonumber\\
&& n_c \simeq  \sqrt{3}/(s \sqrt{\alpha^2-1}) \label{ncestimate}
\end{eqnarray}}
where the ellipsis denote terms higher order in $n$. We note that
$n_c$ here is estimated as $n$ for which the contribution of the
term $\sim n^4$ becomes equal to the leading term $\sim n^2$. We
find that $n_c$ diverges at the transition as $1/\sqrt{\alpha^2-1}$
as one approach the transition from the heating phase; it can also
be made large by tuning either the drive frequency or amplitude
since $s \sim \delta f T$. For $ n\ll n_c$, the behavior $P_n$ is
identical to that on the transition line. A similar result can be
obtained if the transition line is approached from the non-heating
phase. Thus we find that characteristic behavior the return
probability on the transition line can be observed upon approaching
the line below a crossover timescale $n_c$ which diverges at the
transition; $P_n$ has a universal dependence on $s$ for $n \ll n_c$.
This behavior is shown in Fig.\ \ref{fig3} in the high drive
frequency regime $(\hbar \omega_D)/(\pi/L) =100$ where $\delta f =1$
gives the position of the transition line. We clearly see that $\ln
P_n$ remains indistinguishable for $n \le n_c \simeq 50$. The
divergence of $n_c$ at the transition line is shown in the right
panel of Fig.\ \ref{fig3}.

\begin{figure}
\rotatebox{0}{\includegraphics*[width= 0.49 \linewidth]{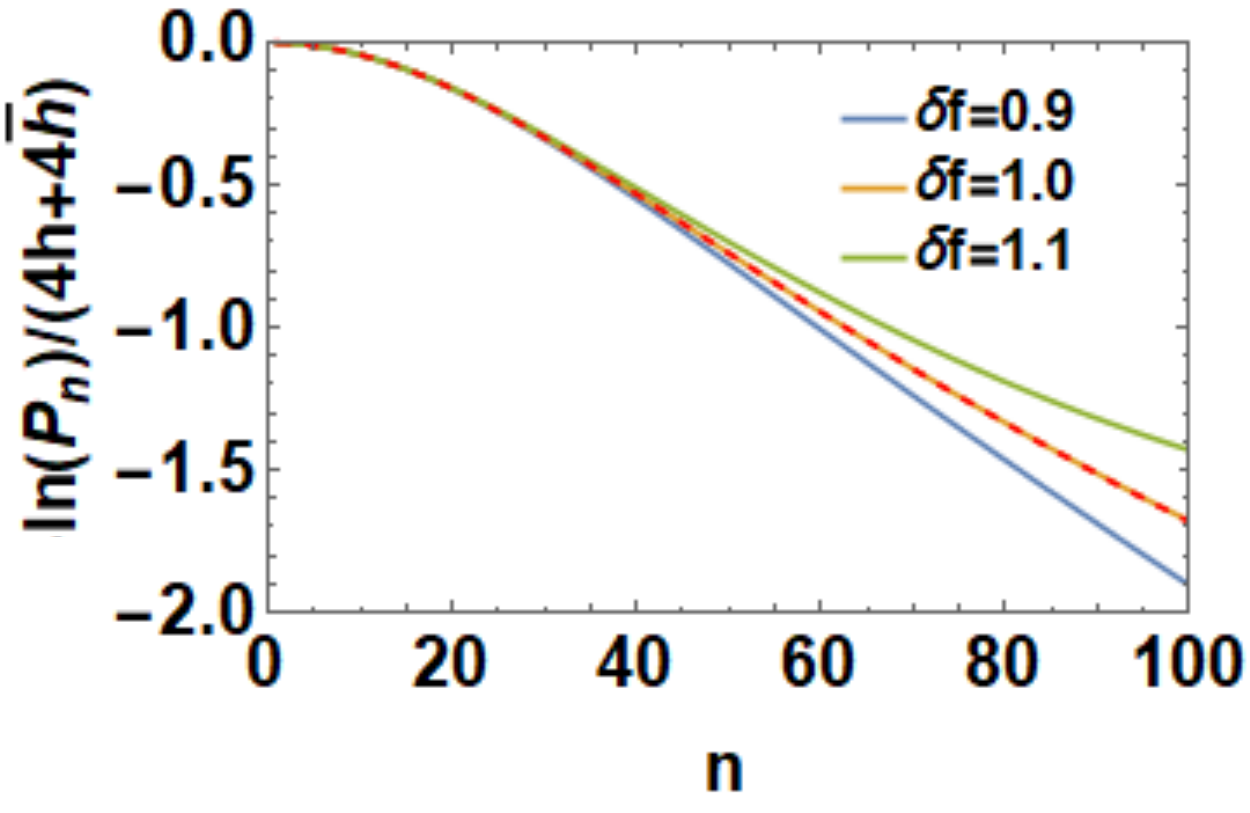}}
\rotatebox{0}{\includegraphics*[width= 0.49 \linewidth]{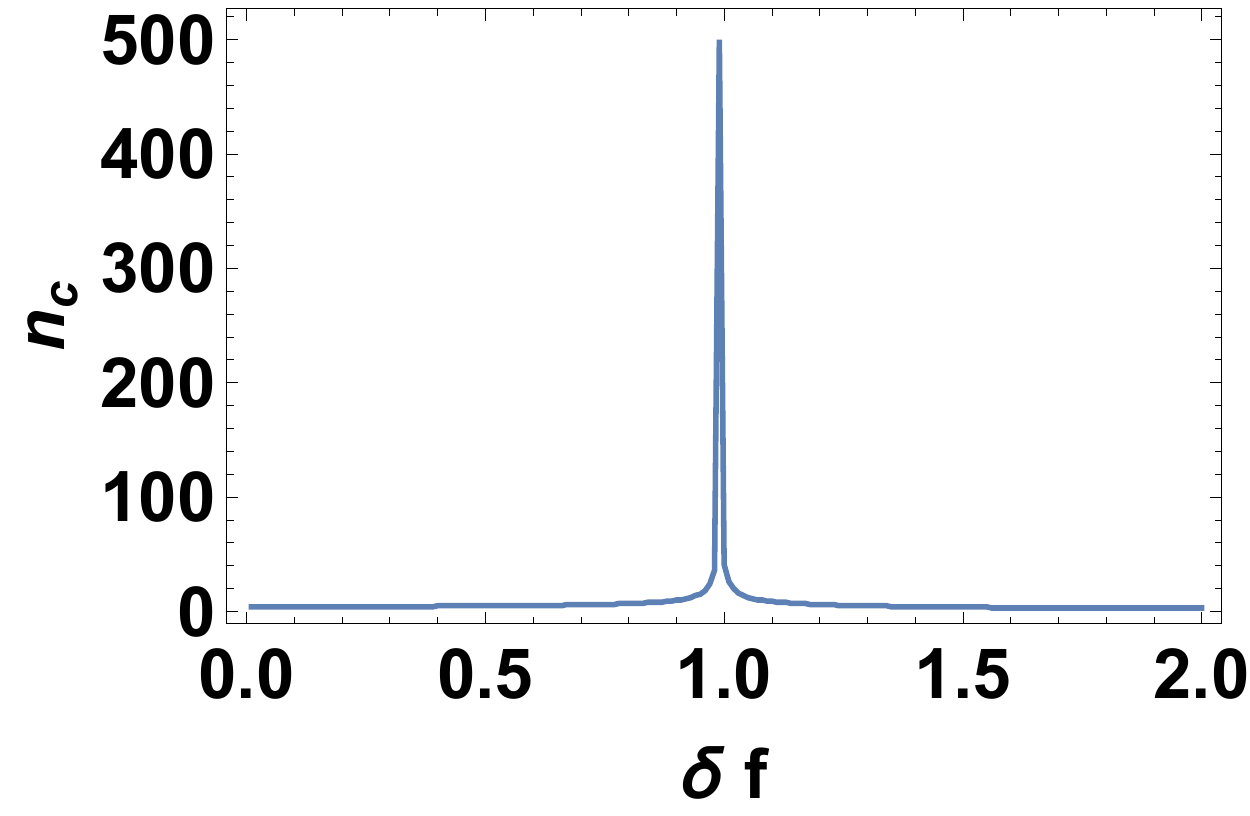}}
\caption{(Color online) Left Panel: Plot of $\ln P_n /(4(h+\bar h))$
as a function of frequency $n$ for $\delta f=1.1$ (non-heating
phase, green line), $1$ (transition line, yellow line) and $0.9$
(heating phase, blue line) with $\omega_D=100$. The red dashed line
is a plot of $-\ln(1+s^2 n^2)$. Right panel: Plot of $n_c$ as a
function of $\delta f$ for $\omega_D=100$ showing a sharp peak at
the transition. For all plots $\pi/L$ has been set to unity, and
$f_0=10$. See text for details.} \label{fig3}
\end{figure}

Next, we study the behavior of the energy density of the driven
system given by
\begin{eqnarray}
E_n &=& \frac{\langle h, \bar h| U^{\dagger n} (T(w) + {\bar T}(\bar
w)) U_n |h,\bar h\rangle}{\langle h,\bar h|h,\bar h\rangle}
\label{enden1}
\end{eqnarray}
To evaluate this, we first consider the holomorphic contribution to
the energy density. To this end, we first move from the cylinder to
the complex plane so that one can write
\begin{eqnarray}
T(w) &=&  \left(\frac{2 \pi z}{L}\right)^2 T(z) + c_1
\label{ttrans1}
\end{eqnarray}
where the constant term $c_1= (2\pi/L)^2 c_0/24$ comes from the
Schwarzian given by Eq.\ \ref{schdef} and $c_0$ denotes the central
charge. In what follows we shall ignore this term since it does not
change upon driving the system. Using this and Eqs.\ \ref{opevol2}
we find
\begin{eqnarray}
&& \langle h| U^{\dagger n} T(w) U^n |h \rangle = \lim_{z_2 \to
\infty, z_1 \to 0} \left(\frac{2 \pi}{L}\right)^{2(1+h)} z_1^h z_2^h
z^2 \nonumber\\
&& \quad \times (\tilde c_n z+\tilde d_n)^{-4} \langle 0| \phi(z_2)
T(z_n) \phi(z_1) |0 \rangle \label{ttrans2}
\end{eqnarray}
where we have used $w= i x$ so that $z= \exp[2\pi i x/L]$ and $w_i=
\tau_i$ for $i=1,2$ with $\tau_1 \to -\infty$, $\tau_2 \to \infty$
leading to the limits $z_2 \to \infty$ and $z_1 \to 0$.

Eq.\ \ref{ttrans2} reduces the computation of the energy density to
that of computation of $\langle \phi T \phi \rangle$ correlator in
the CFT vacuum state. This can be done in a straightforward manner
using standard identity \cite{bpz1} and yields
\begin{widetext}
\begin{eqnarray}
\langle 0| \phi(z_2) T(z_n) \phi(z_1) |0 \rangle &=& h \left[
(z_n-z_2)^{-2}(z_2-z_1)^{-2h} + (z_n-z_1)^{-2}(z_2-z_1)^{-2h}
\right.
\nonumber\\
&& \left. -2 (z_n-z_2)^{-1}(z_2-z_1)^{-(2h+1)} - 2(z_n-z_1)^{-1}
(z_2-z_1)^{-(2h+1)} \right] \label{ttrans3}
\end{eqnarray}
\end{widetext}
We substitute Eq.\ \ref{ttrans3} in Eq.\ \ref{ttrans2} and find that
in the limit $z_2 \to \infty$ only the second term in the right side
of Eq. \ \ref{ttrans3} contribute. Taking the $z_1 \to 0$ limit, we
finally obtain the holomorphic contribution to the energy density to
be
\begin{eqnarray}
E_n^{\rm hol} &=& \left(\frac{2\pi}{L}\right)^2
\frac{h}{(R_{1n}(x) R_{2n}(x))^2} \label{holen}\\
R_{1(2) n}(x) &=& [\tilde a_n (\tilde c_n) e^{\pi i x/L} + \tilde
b_n(\tilde d_n) e^{-\pi i x/L}]
\end{eqnarray}
The antiholomorphic contribution can be simply read off from this to
be $E_n^{\rm hol}(x) = E_n^{\rm ant-hol}(-x)$. Thus the total energy
density which is the sum of the holomorphic and anti-holomorphic
parts are given by
\begin{eqnarray}
E_n(x) &=& \left(\frac{2\pi}{L}\right)^2 h \left( [R_{1n}(x) R_{2n}(x)]^{-2} \right. \nonumber\\
&& \left. + [R_{1n}(-x) R_{2n}(-x)]^{-2} \right) \label{enden2}
\end{eqnarray}
We note that the expressions of the energy density of the asymptotic
states is identical to that obtained for the vacuum state in the
strip geometry in Ref.\ \onlinecite{cft4}. The only difference
appears in the prefactor; the asymptotic states have a prefactor $h$
while the vacuum in the strip geometry has a prefactor of $c_0/32$.
Thus we expect similar emergent spatial structure for $E_n(x)$ as
discussed in Ref.\ \onlinecite{cft4}. Below, we analyze this
phenomenon for the continuous drive protocol. To this end, we
analytically continue to real time, substitute Eq.\ \ref{numexp2} in
Eq.\ \ref{enden2} and, using Eq.\ \ref{comp1}, obtain for the
heating phase
\begin{eqnarray}
E_n(x) &=& \left(\frac{2\pi}{L}\right)^2  2 h (\alpha^2-1)^2
\frac{Q_1^2(x)+Q_2^2(x)}{(Q_1^2(x)-Q_2^2(x))^2} \nonumber\\
Q_1(x) &=& \alpha \cos\left(2 \pi x/L \right)
(\cosh[2n\theta]-1)\nonumber\\
&& + \alpha^2 \cosh[2 n \theta] -1 \nonumber\\
Q_2(x) &=& \alpha \sqrt{\alpha^2-1} \sin \left(2 \pi x/L \right)
\sinh[2 n \theta] \label{heatingen}
\end{eqnarray}

\begin{figure}
\rotatebox{0}{\includegraphics*[width= 0.48 \linewidth]{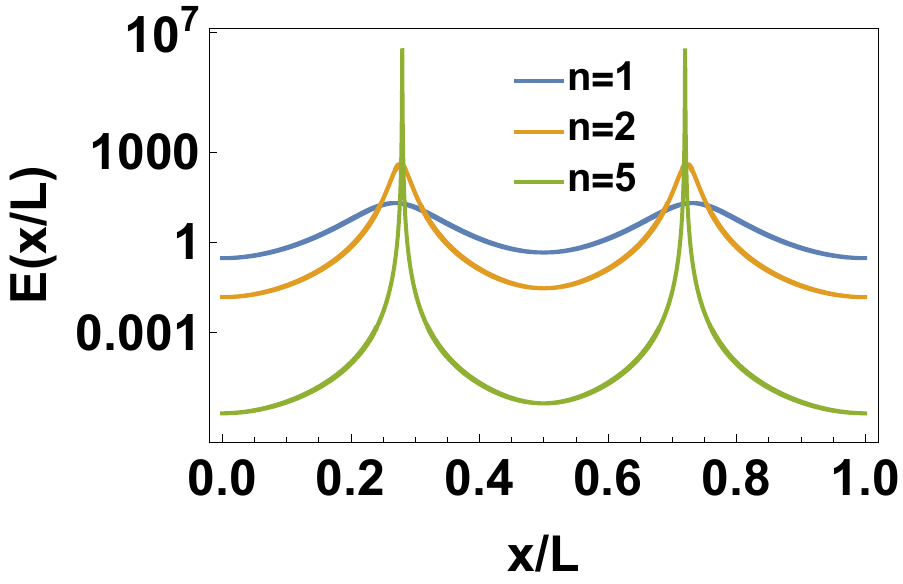}}
\rotatebox{0}{\includegraphics*[width= 0.48 \linewidth]{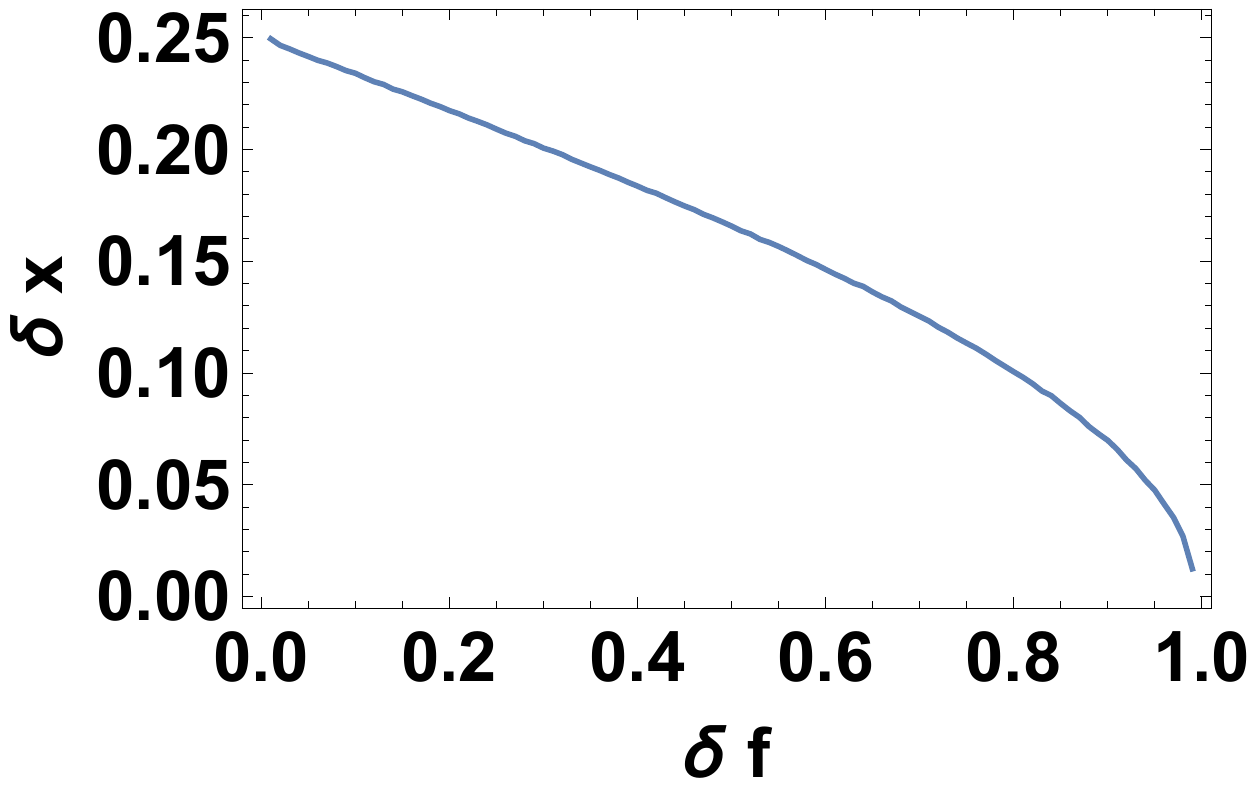}}
\caption{Left Panel: Plot of $E_n(x)$ as a function of $x$ for
several representative values of $n$ in the heating phase with
$\omega_D=3$ and$\delta f=0.04$. Right panel: Plot of $\delta x/L$,
as a function of $\delta f$ for $\omega_D=100$. The peaks of
$E_n(x)$ occur at $L/2 \pm \delta x$. For all plots $\pi/L$ is set
to unity and $f_0=10$. See text for details.} \label{fig4}
\end{figure}

We note from Eq.\ \ref{heatingen} that for a generic $x$, $E_n(x)$
decays exponentially with $n$ with the large $n$ limit: $E_n(x) \sim
e^{-4 n \theta}$. This behavior can be in Fig.\ \ref{fig4} where
$E_n(x)$ is plotted for several $n$ as a function of $x$. The plot
shows that $E_n(x)$, at large $n$, decays at all positions except at
two places for which the leading terms of $Q_1^2(x)$ and $Q_2^2(x)$
in the large $n$ limit cancels each other. A straightforward
calculation shows that
\begin{eqnarray}
x_c^{\pm} &=&  \frac{L}{2} \pm \delta x,\quad  \delta x =
\frac{L}{2\pi} \arccos(1/\alpha) \label{peakp}
\end{eqnarray}
within first order FPT. Thus the position of the peaks of $E_n(x)$
in the large $n$ limit moves from $L/2$ to $L/4$ and $3L/4$ as one
moves from the phase boundary to deep inside the heating phase. This
behavior is supported for exact numerics as can be seen in the right
panel of Fig.\ \ref{fig4} where we plot the peak positions ($\pm
\delta x/L$) as a function of $\delta f$ for $\hbar \omega_D=100
(\pi/L)$.

\begin{figure}
\rotatebox{0}{\includegraphics*[width= 0.48 \linewidth]{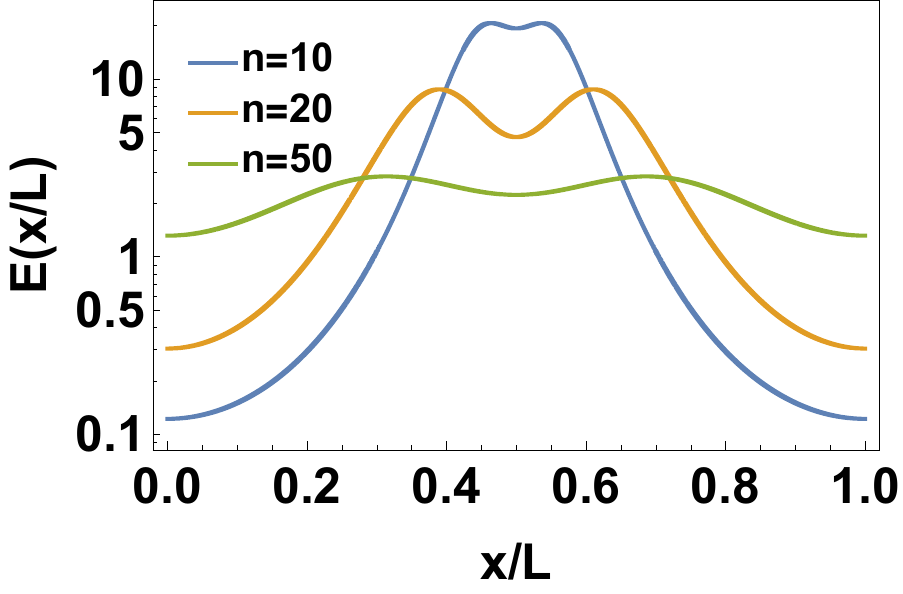}}
\rotatebox{0}{\includegraphics*[width= 0.48 \linewidth]{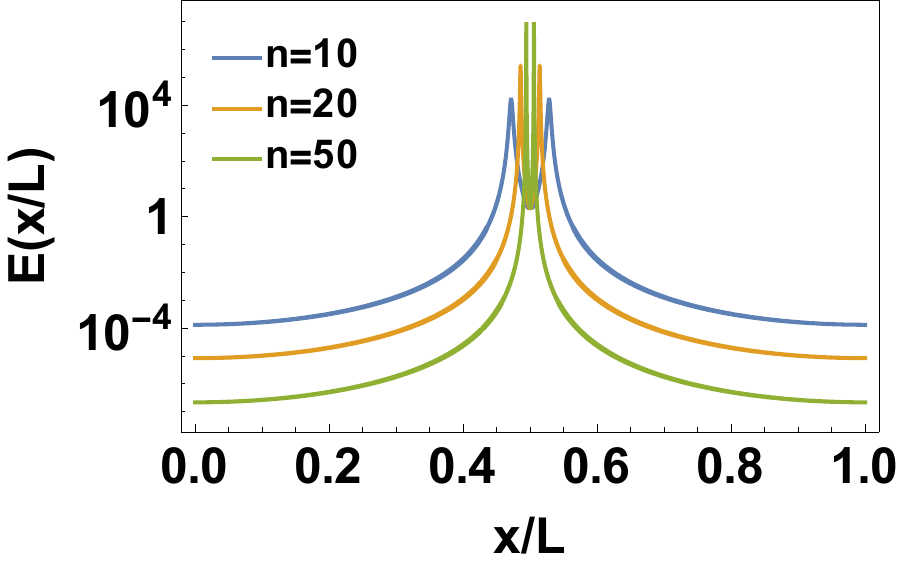}}
\caption{Left Panel: Plot of $E_n(x)$ as a function of $x$ for
several representative values of $n$ in the non-heating phase ith
$\omega_D=3$ and$\delta f=0.4$. Right panel: Same for $\omega_D=3$
and $\delta f=0.27$ when the system is almost on the transition line
showing the development of a single peak at $x=L/2$ for large $n$
For all plots $\pi/L$ is set to unity and $f_0=10$. See text for
details.} \label{fig5}
\end{figure}

In contrast for the non-heating phase we find using Eq.\
\ref{numexp3}, $E_n(x)$, in real time, is given by Eq.\
\ref{heatingen} with $Q_1(x) \to Q'_1(x)$ and $Q_2(x) \to Q'_2(x)$,
where
\begin{eqnarray}
Q'_1(x) &=& \alpha \cos\left(2 \pi x/L \right) (1-\cos[2n\theta]) +
1-\alpha^2 \cos[2 n \theta] \nonumber\\
Q'_2(x) &=& \alpha \sqrt{\alpha^2-1} \sin \left(2 \pi x/L \right)
\sin[2 n \theta] \label{nonheatingen}
\end{eqnarray}
Thus $E_n(x)$ is an oscillatory function of $n$ as shown in the left
panel of Fig.\ \ref{fig5} where $E_n(x)$ is plotted as a function of
$x$ of several $n$.

Finally on the transition line, using Eq.\ \ref{numexp4}, we find
\begin{eqnarray}
E_n(x) &=& \left(\frac{2\pi}{L}\right)^2  h \frac{2(Q"_1^2(x) +
Q"_2^2(x))}{(Q"_1^2(x) -
Q"_2^2(x))^2} \nonumber\\
Q"_1(x) &=& 1 + 2 n^2 s^2 (1+\cos(2 \pi x/L)) \nonumber\\
Q"_2(x) &=& 2sn \sin(2\pi x/L) \label{tranen}
\end{eqnarray}
where we have analytically continued to real time. Thus we find the
peak of $E_n(x)$ in the large $n$ limit occurs at $x=L/2$ for which
$Q"_1(x)=1$ and $Q"_2(x)=0$. In contrast, for $x=0,L$, we have
$Q"_1(x)= 1+4n^2 s^2$ and $Q"_2(x)=0$, so that $E_n(x) \sim 1/(1+ 4
n^2 s^2)^2$ for all $n$. This behavior is consistent with the fact
that $n'_c$ diverges at the transition. A plot of $E_n(x)$ as a
function of $x$ for several $n$, shown in the right panel of Fig.\
\ref{fig5}, confirms this behavior.

\begin{figure}
\rotatebox{0}{\includegraphics*[width= 0.49 \linewidth]{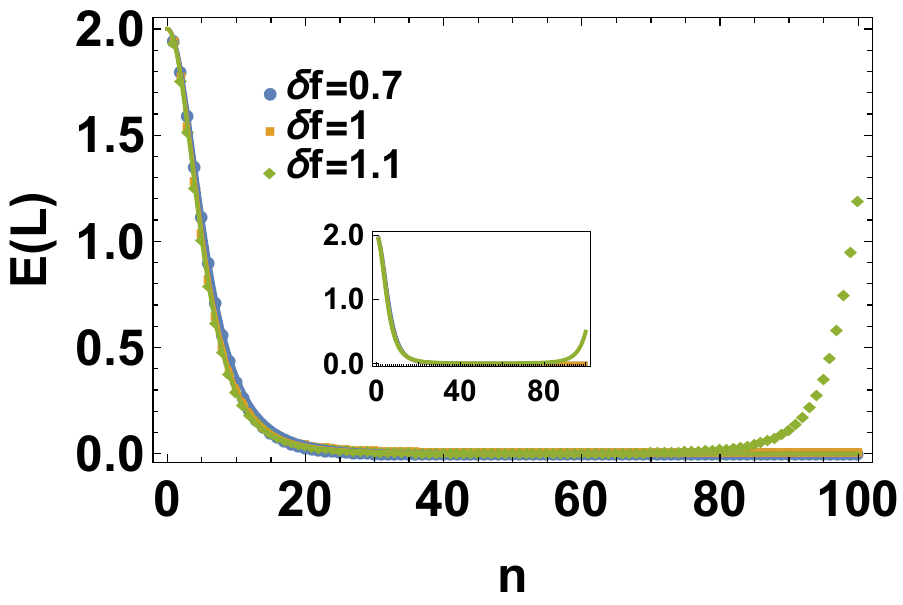}}
\rotatebox{0}{\includegraphics*[width= 0.49 \linewidth]{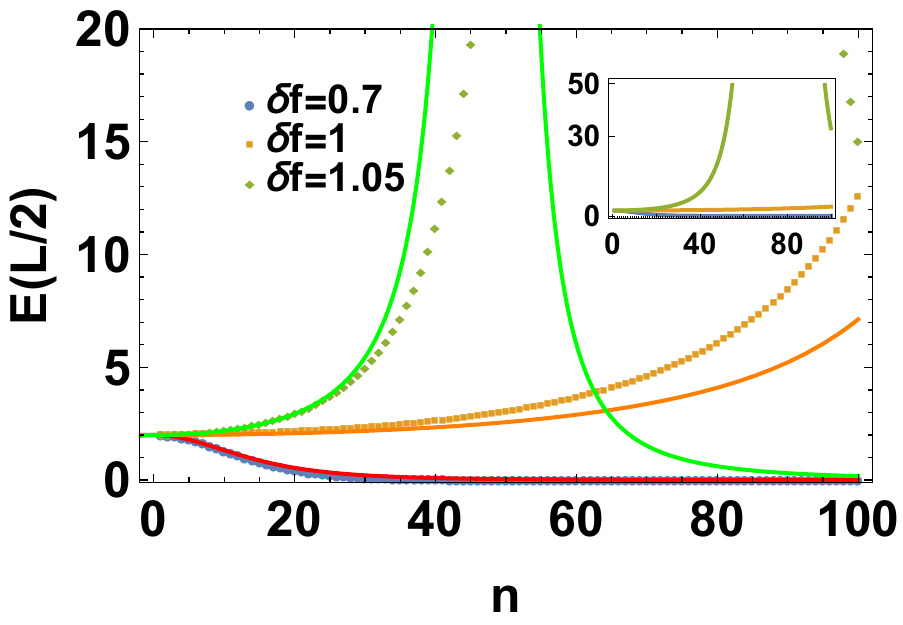}}
\caption{(Color online) Left Panel: Plot of $E_n(L)$ as a function
of $n$ in the heating ($\delta f=0.7$, blue points), non-heating
($\delta f=1.1$, green points) phase, and on the transition line
($\delta f=1$, orange points). The solid lines shows the scaling
law$1/(1+\mu n^2)^2$. The inset shows results obtained using first
order FPT. Right panel: Plot of $E_n(L/2)$ as a function of $n$
showing different behavior for $E_n(L/2)$ in the heating phase
($\delta f=0.7$, blue points), non-heating phase ($\delta f=1.05$,
yellow points) and on the transition line ($\delta f=1$, yellow
points). The solid lines shows the scaling law $1/(1+\mu^{\prime}
n^2)^2$ for each case. For the heating phase the color of the solid
line is red for enhanced visibility. Here we have used $f_0=10$
$\omega_D=100$ and $\pi/L$ has been set to unity for all plots. See
text for details.} \label{fig6}
\end{figure}

We also note that $E_n(x)$ takes particulary simple forms at the
center and end of the chains where it is given by
\begin{eqnarray}
E_n(0)&=& E_n(L) = \left(\frac{2\pi}{L}\right)^2  \frac{
2(\alpha-1)^2 h}{(\alpha \cosh[2 n
\theta]-1)^2}   \nonumber\\
E_n(L/2) &=& \left(\frac{2\pi}{L}\right)^2
 \frac{2(\alpha+1)^2 h}{(\alpha \cosh[2 n \theta]+1)^2} \label{endhalf}
\end{eqnarray}
From Eq.\ \ref{endhalf}, we once again find the existence of a
crossover timescale $n'_c = n_c \sqrt{(1+ \alpha)/(1+ 4 \alpha)}$
below which $E_n(0)$ decays as $1/(1+ \mu n^2)^2$ where {$\mu= 2 s^2 \alpha(1+ \alpha)$}. 
This behavior is verified in
the left panel of Fig.\ \ref{fig6} which shows the behavior of
$E_n(0)$ (with $E_n(0)=E_n(L)$) as a function of $n$; we find that
the curves for the heating and non-heating phase become identical to
the transition line for $n \ne n'_c$. For $n \gg n'_c$, the decay
becomes exponential in the heating phase. In contrast, for
$E_n(L/2)$, plotted in the right panel of Fig.\ \ref{fig6}, there is
a clear distinction between behavior of $E_n$ in the heating phase
and on the transition line. This can be attributed to the fact that
$x=L/2$ coincides with the position of the peak on the transition
line, while $E_n(L/2)$ decays in the heating phase.

\subsection{Correlation functions}
\label{corrfns}

In this subsection, we present results for both equal-time and
unequal-time correlation functions of primary operators and the
stress tensor.

\subsubsection{Equal time correlation function}
\label{etc}

We begin with the analysis of the equal-time correlation function
starting from the asymptotic state $|h, \bar h\rangle$ given by
\begin{eqnarray}
C_n(x_1,x_2) &=& \frac{ \langle h,\bar h|U^{\dagger n} \phi(w_1,\bar
w_1) \phi(w_2,\bar w_2) U^n |h,\bar h\rangle}{\langle h,\bar
h|h,\bar h\rangle}  \nonumber\\ \label{etcf1}
\end{eqnarray}
where $w_j (\bar w_j) = +(-) i x_j$ for $j=1,2$.

We first compute the holomorphic part of $C_n(x_1,x_2)$ which is
given by
\begin{eqnarray}
&& C_n^{\rm hol}(x_1,x_2) = \frac{ \langle h|U^n \phi(w_1)
\phi(w_2) U^n |h,\rangle }{\langle h | h \rangle} \nonumber\\
&=& \lim_{z_3 \to \infty, z_4 \to 0} \prod_{j=1,2}
\left(\frac{\partial w_j}{\partial z_j}\right)^{-h} \prod_{j=1,2}
\left(\frac{\partial z_{jn}}{\partial z_j}\right)^{h}
\nonumber\\
&& \times z_3^{2h} {\langle 0| \phi(z_3) \phi(z_{n1}) \phi(z_{n2})
\phi(z_4) |0 \rangle } \label{etcft2}
\end{eqnarray}
where $z_{jn} = (\tilde a_n z_j +\tilde b_n)/(\tilde c_n z_j +
\tilde d_n)$. Here we choose the dimension of the primary fields
$\phi$ to be same that of the state $|h\rangle$. The
anti-holomorphic part an be written similarly in terms of $\bar z$
and $\bar z_{jn}$. To compute the four-point correlators of the
primary fields, we use the standard result which expresses these in
terms of the cross ratio of the complex coordinates $z_i$. To this
end, we define $z_{ij} = z_i -z_j$ with $z_k= z_{k n}$ for $k=1,2$.
Using this notation, we define the cross ratio
\begin{eqnarray}
\eta &=& \frac{z_{31} z_{24}}{z_{32} z_{14}} =
\frac{(z_3-z_{1n})(z_{2n}-z_4)}{(z_3-z_{2n})(z_{1n}-z_4)}
\label{creq1}
\end{eqnarray}
We note that for $z_{3} \to \infty$ and $z_4\to 0$, we have
\begin{eqnarray}
&& \eta \to  y_n(x_1,x_2) = \frac{z_{2n}}{z_{1n}} \label{creq2} \\
&& = \frac{(\tilde a_n e^{i \pi x_2/L} +\tilde b_n e^{-i \pi x_2/L})
(\tilde c_n e^{i \pi x_1/L} + \tilde d_n e^{-i \pi x_1 /L})}{(\tilde
a_n e^{ i \pi x_1 /L}+\tilde b_n e^{-i \pi x_1/L}) (\tilde c_n e^{i
\pi x_2/L} + \tilde d_n e^{- i \pi x_2/L})} \nonumber \label{creq2}
\end{eqnarray}
A similar result for $\bar y_n$ can be obtained for the
anti-holomorphic part by replacing $x_i \to -x_i$.

To compute the four-point correlator, we first define the quantity
\begin{eqnarray}
{\mathcal C}_n  &=& \lim_{z, {\bar z} \to \infty} z^{2h} {\bar z}^{2
\bar h} \langle 0| \phi(z,\bar z) \phi(z_{1n},\bar z_{1n})
\nonumber\\
&& \times \phi(z_{2n}, \bar z_{2n}) \phi(0,0) |0\rangle
\label{corr1}
\end{eqnarray}
To evaluate this, we first note that in the non-heating phase and on
the transition line, $|1-z_{2n}/z_{1n}|,  |1-\bar z_{2n}/\bar
z_{1n}|\to 0$ for large $n$ in Euclidean time. To take advantage of
this limit, we make the conformal transformation (where $z_1 \equiv
z_{1n}$ and $z_2 \equiv z_{2n}$)
\begin{eqnarray}
z_i \to z'_i = -\frac{z (z_i-z_{2n})}{(z_{1n}-z_i) z_{2n}}
\label{contrans1}
\end{eqnarray}
so that in the new coordinate $z'_{2n} =0$.  A similar
transformation is done for $\bar z_{i}$. Then using the standard
transformation rule of operators $\phi(z,\bar z) \to (\partial
z'_i/\partial z_i)^{-h} (\partial z'_i/\partial z_i)^{-h}
\phi(z'_i)$, one can write, after some straightforward algebra
\cite{cftlit}
\begin{eqnarray}
{\mathcal C}_n &=& z_{1n}^{-2h} \bar z_{1n}^{-2 \bar h} {\mathcal
F}(1-y_n; 1-\bar y_n) \label{correq2}
\end{eqnarray}
where $y_n = z_{2n}/z_{1n}$ and $\bar y_n = \bar z_{2n}/\bar
z_{1n}$. The advantage of using this form for ${\mathcal F}$ which
admits conformal block decomposition is that one can write down a
perturbative expansion of the blocks around $|1-y_n|, |1-\bar
y_n|=0$. This allows us to obtain an analytic, albeit perturbative
expression of $C_n(x_1,x_2)$ for arbitrary $h, \bar h$. For the
driven problem, $|1-y_n|, |1-\bar y_n| \ll 0$ (in Euclidean time) in
the non-heating phase  and on the transition line for all $x_1, x_2$
in the large $n$ limit. Also, for all phases, this limit holds for
all $n$ only for $|x_1 -x_2|/L\ll 1$. The perturbative results that
we chart out next is expected to be accurate in these limits. For
the present case, one obtains \cite{cftlit2}
\begin{widetext}
\begin{eqnarray}
{\mathcal F} &=& \sum_p C_{hhp}^2 \mathcal{V}_p(y_n,h)
\bar {\mathcal V}_{\bar p} ( \bar y_n, \bar h) \nonumber\\
\mathcal{V}_p(y_n,h)&=& (1-y_n)^{h_p-2h} \sum_k F_k (1-y_{n})^k,
\quad  \bar {\mathcal V}_{\bar p}(\bar y_n,\bar h) =  (1-\bar
y_n)^{\bar h_p-2\bar h} \sum_k F_k
(1-\bar y_{n})^k \nonumber\\
\sum_k F_k x^k &=& 1 + \frac{h_p}{2} x + \frac{h_p \left(h_p
\left(h_p \left(c_0+8 h_p+8\right)+2 (c_0+4 h-4)\right)+c_0+8 h (2
h-1)\right)+8 h^2}{8 h_p \left(c_0 +8 h_p-5\right)+4 c_0} x^2 +
{\mathcal{O}}(x^3) \label{cftres1}
\end{eqnarray}
\end{widetext}
where $\sum_p$ denotes sum over primaries and $\mathcal{V}_p$
denotes the $p$-th conformal block with dimensions $h_p, \bar h_p$
and the coefficients $C_{hhp}$ that depend on the details of the
CFT. We note here that the identity block corresponds to $h_p=0$ and
for this block $C_{hhI}=1$.

Substituting Eq.\ \ref{cftres1} in Eq.\ \ref{etcft2} (and its
corresponding anti-holomorphic part), one obtains
\begin{eqnarray}
\frac{C_n(x_1,x_2)}{C_0(x_1,x_2)}  &=& \sum_p C_{hhp}^2 (1-
y_n)^{h_p}( 1- \bar y_n)^{\bar h_p} \label{corrpert1} \\
C_0(x_1,x_2) &=& \left(\frac{2\pi}{L}\right)^{2(h+\bar h)}
\frac{(z_1 z_2)^h }{(z_1-z_2)^{2h}} \frac{(\bar z_1 \bar z_2
)^{\bar{h}}}{(\bar z_1-\bar z_2)^{2\bar h}} \nonumber
\end{eqnarray}
where only the leading term of the $\sum_k F_k x^k \simeq 1$ is
retained.

\begin{figure}
\rotatebox{0}{\includegraphics*[width= 0.49 \linewidth]{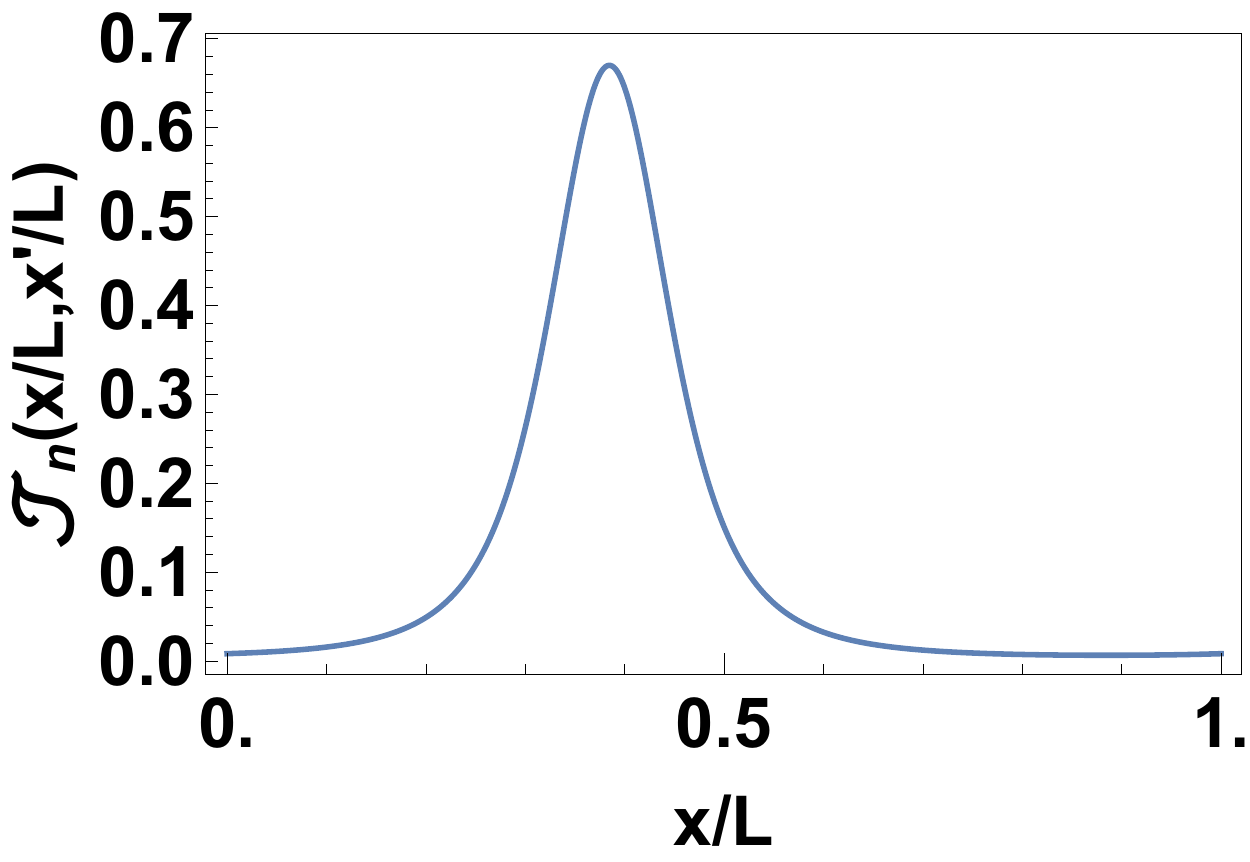}}
\rotatebox{0}{\includegraphics*[width= 0.49 \linewidth]{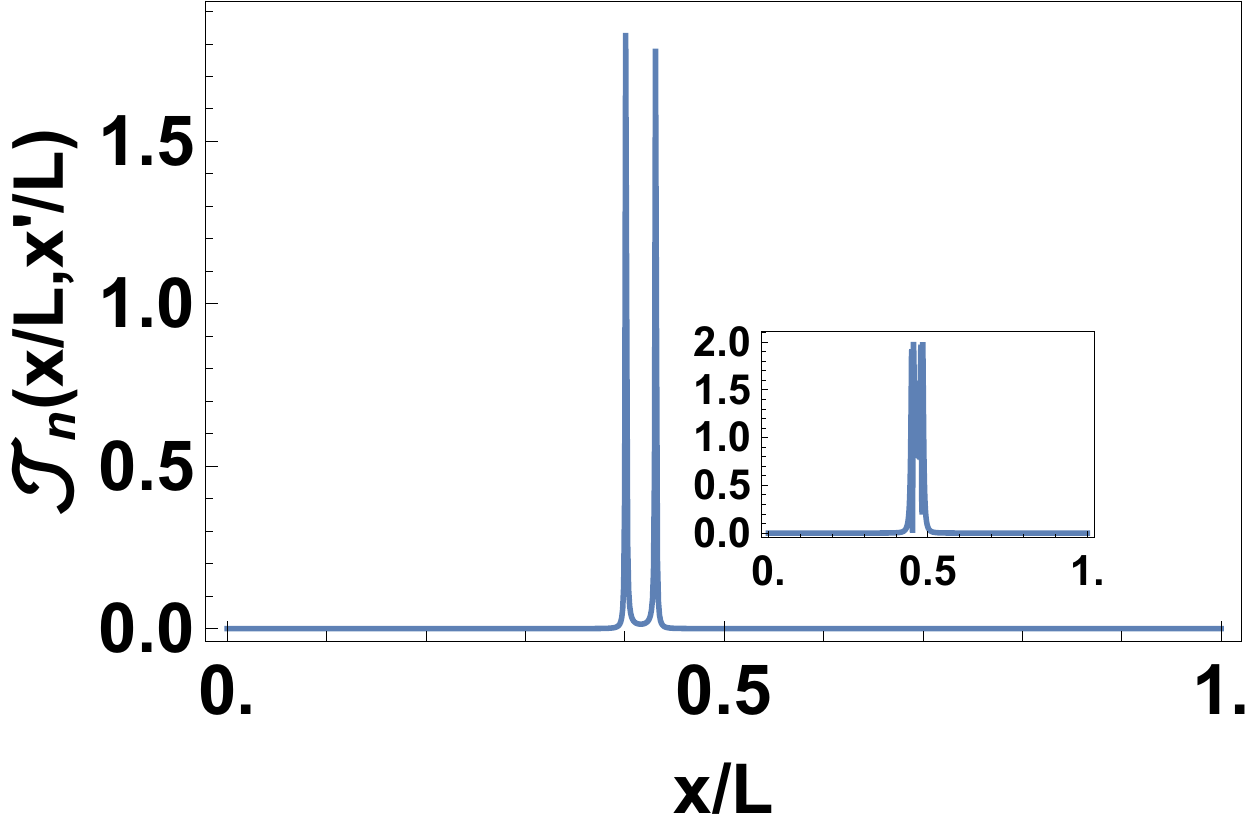}}
\rotatebox{0}{\includegraphics*[width= 0.49 \linewidth]{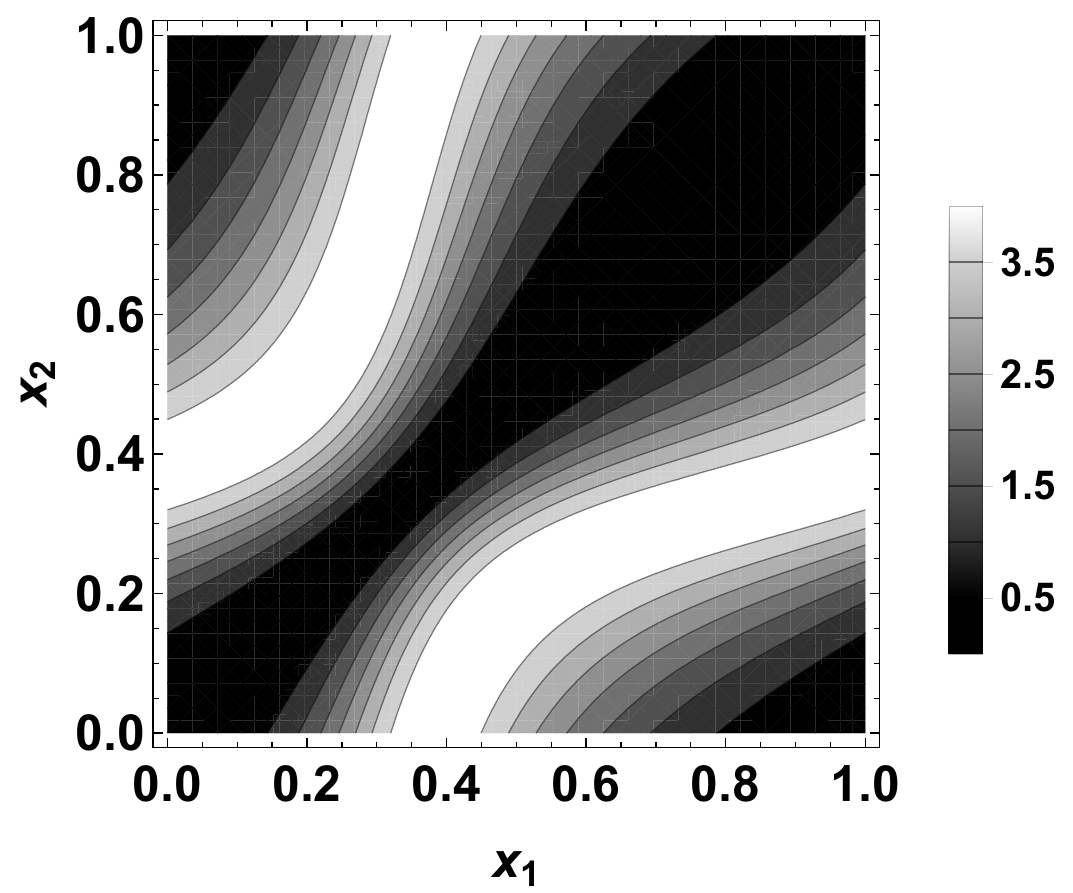}}
\rotatebox{0}{\includegraphics*[width= 0.49 \linewidth]{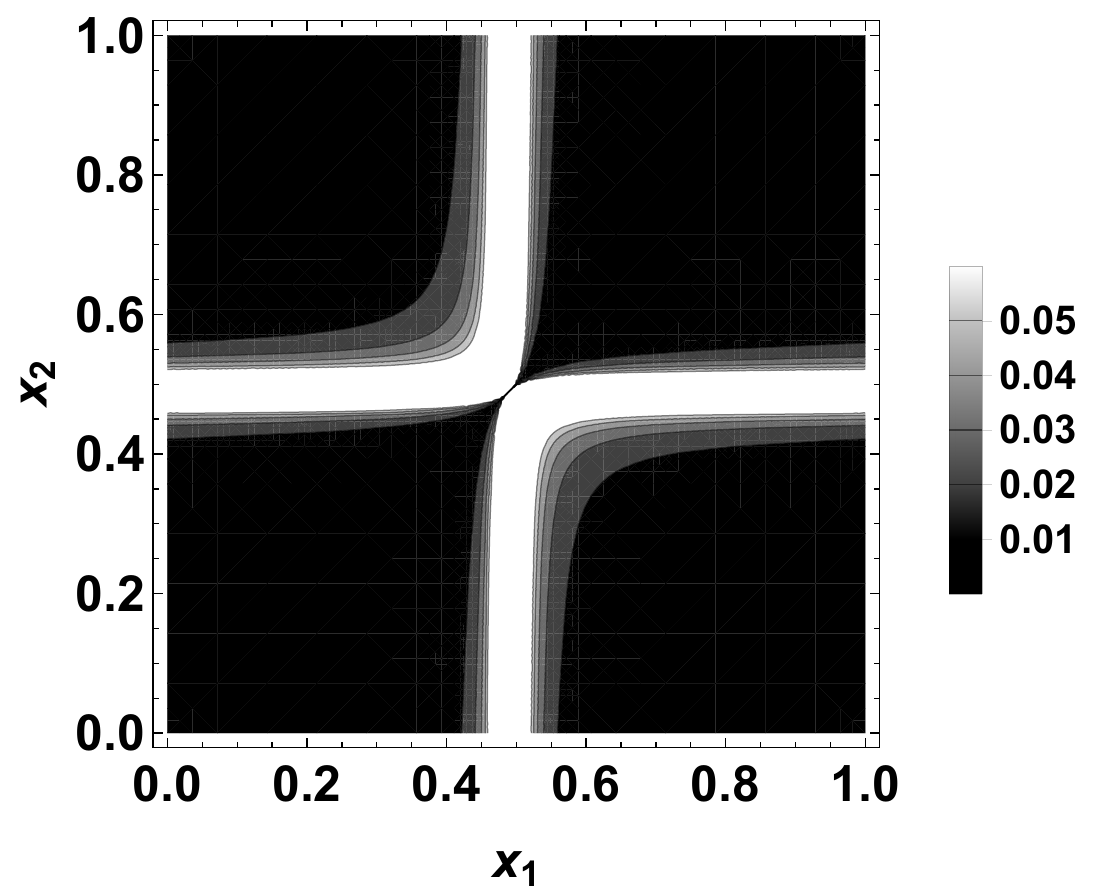}}
\caption{(Color online) Top left Panel: Plot of {
${\mathcal J}_n(x_1,x_2)= (C_n(x_1,x_2)_{\rm
univ}/C_0(x_1,x_2)-1)/(2h^2/c_0)$ as a function of $x_1/L \equiv
x/L$} in the non-heating phase ($\delta f=1.5$, $\omega_D=100$) for
$x'/L= x/L + 0.03$ and $n=100$. Top Right panel: Similar plot for
the heating phase ($\delta f=1.1$). The inset shows the behavior of
the correlation function on the transition line ($\delta f =1$).
Bottom left panel: Plot of ${\mathcal J}_n(x_1,x_2)$ as a function
of $x_1$ and $x_2$ in the non-heating phase for $n=100$ showing
emergence of spatial structure. Bottom right panel: Similar to the
top left panel but for $\delta f=1$ where the system is on the
transition line. Here we have used $f_0=10$ $\omega_D=100$ and
$\pi/L$ has been set to unity for all plots. See text for details.}
\label{fig10}
\end{figure}

Next, we discuss the contribution of the identity block which is
universal since $C_{hhI}=1 = C_{\bar h \bar h I}$. For this block,
$h_p=\bar h_p=0$. We note that if we retain only the first order
term, we find from Eq.\ \ref{corrpert1} that $C_n(x_1,x_2)$ becomes
independent of $n$. The first non-trivial contribution of the
identity block arises from the $x^2$ term in the expansion of
$\sum_k F_k x^k$. This universal contribution is given by
\begin{eqnarray}
\frac{C_n(x_1,x_2)_{\rm univ}}{C_0(x_1,x_2)}  &=& 1+
\frac{2h^2}{c_0} {\rm Re}[(1- y_n)^{2}] \label{corrpert2}
\end{eqnarray}
Thus in this limit, the identity block contribution to the deviation
of $C_n$ from its equilibrium value provides a measure of the
central charge. A plot of ${\mathcal J}_n(x_1,x_2)= (
C_n(x_1,x_2)_{\rm univ}/C_0(x_1,x_2) -1)/(2h^2/c_0)$ (after analytic
continuation to real time) is shown in the top panels of Fig.\
\ref{fig10} as a function $x_1/L= x/L$ for $x_2/L=x_1/L +0.03$ after
$n=100$ cycles of the drive. For these plots we have chosen $\hbar
\omega_D = 100 \pi/L$. The top left panel shows the behavior of
$C_n(x_1,x_2)_{\rm univ}$ in the non-heating phase ($\delta f=1.5$)
displaying a broad oscillatory structure. In contrast, in the top
right panel, for the heating phase ($\delta f=0.9$) it displays two
sharp peaks consistent with the behavior of $E_n(x)$. The position
of these peaks shift to $L/2$ on the transition line as can be seen
from the inset of top right panel of Fig.\ \ref{fig10}. The bottom
panels show the behavior of $C_n(x_1,x_2)_{\rm univ}$ for large
$n=100$ in the non-heating phase (left panel) and on the transition
line (right panel) where the perturbative expansion of ${\mathcal
F}$ is expected to be accurate. We find clear emergence of spatial
pattern in the non-heating phase in contrast to the behavior of
$E_n$. We shall discuss this behavior in more details in the context
of unequal-time correlation function.

In the large central charge $c_0$ limit, with the conformal
dimensions held fixed, the Virasoro conformal blocks reduce to
global conformal blocks which are given in terms of hypergeometric
functions, by \cite{cftlit6},
\begin{eqnarray}
\mathcal{V}_p(x) &=& x^{h_p-2h}\,{}_2F_1\left( h_p , h_p;2h_p;
x\right) + {\mathcal{O}}(1/c).
\end{eqnarray}
Hence the correlator is given by
\begin{eqnarray}
&& \frac{C_n(x_1,x_2)}{C_0(x_1,x_2)} \simeq  \sum_{p\neq \mathbb{I}}
C_{hhp}^2 (1- y_n)^{h_p}( 1- \bar y_n)^{\bar h_p} \label{corrC}
\\
&& \times {}_2F_1\left( h_p , h_p;2h_p; 1-y_n\right){}_2F_1\left(
\bar h_p , \bar h_p;2\bar h_p; 1-\bar y_n\right). \nonumber
\end{eqnarray}
where $C_0(x_1,x_2)$ is defined in Eq.\ \ref{corrpert1}. Note that
since the Virasoro blocks reduce to global block in this limit,
there is no identity block, i.e., $h_p \neq 0$.

Finally, we note for CFTs with large central charge $c_0 \gg 1$,
when the asymptotic states have large conformal dimension $H \gg h$,
with $h/c_0$ and $H/c_0$ both held fixed, a closed form answer is
available via the monodromy methods. Therefore we compute the
equal-time correlation function $C'_n(x_1,x_2)= \langle H, \bar H|
\phi(w_1,\bar w_1) \phi(w_2, \bar w_2) |H, \bar H\rangle/\langle
H,\bar H|H, \bar H\rangle$. This can be done exactly in the same way
as charted out above; the only difference is that one needs to keep
track of two operator dimension $H$ and $h$. A straightforward
calculation shows that in this case one has \cite{cftlit3}
\begin{eqnarray}
{\mathcal V}_p(z_{1n},z_{2n},h) &=& \left(\frac{ a_0
(z_{1n}z_{2n})^{(a_0-1)/2}}{z_{1n}^{a_0}-z_{2n}^{a_0}} \right)^{2h}
\nonumber\\
&& \times \left( \frac{4(1-y_n^{a_0/2})}{a_0(1+y_n^{a_0/2})}
\right)^{h_p} \label{cftlargec}
\end{eqnarray}
with $a_0= \sqrt{1-24 H/c_0}$. A similar expression can be obtained
for $\bar {\mathcal V}_{\bar p}$ by substituting $ z_{jn} \to \bar
z_{jn}$ and $ h, h_p \to \bar h, \bar h_p$. One can then write $C_n$
in terms of ${\mathcal V}_p$ and $\bar {\mathcal V}_p$ as
\begin{eqnarray}
C'_n(x_1,x_2) &=& \left(\frac{2 \pi }{L}\right)^{2(h+ \bar h)}
\sum_p
C^2_{hhp} {\mathcal V}_p(z_{1n},z_{2n},h)  \nonumber\\
&& \times \bar {\mathcal V}_{\bar p}(\bar z_{1n},\bar z_{2n},\bar
h)\left( { z_1 z_2 }\right)^{h}\left( { \bar z_1 \bar z_2
}\right)^{\bar h} . \label{cftres2}
\end{eqnarray}
A motivation for studying such large $c_0$ CFTs comes from the
AdS/CFT correspondence. These are CFTs which are expected to have
semiclassical gravity duals. The global block answer for light
correlators is reproduced in bulk AdS by the geodesic Witten
diagrams \cite{cftlit7}. The Virasoro vacuum block is non-trivial in
two dimensions unlike its higher dimensional versions as it contains
the stress tensor and its descendants. {In the large
$c_0$ limit}, the dynamics of the Virasoro vacuum block matches with
results from semiclassical gravity \cite{cftlit3}. In CFT${}_2$,
since one has Virasoro, one can use the geodesic Witten diagrams to
interpolate between the global and the semiclassical monodromy
answer (when pair of operator conformal dimensions scale with the
central charge) by taking into account backreaction due to the heavy
geodesics\cite{cftlit8}. The time-dependent drive of an
inhomogeneous metric will have implications for the physics of black
holes in the dual gravitational theory. We are not going to explore
this issue further here.

\subsubsection{Unequal time correlation function}
\label{utc}

In this subsection, we compute the unequal-time correlation function
of the primary fields $G_n(x_1,x_2)$. We note that unequal-time
correlation functions of the vacuum state, unlike-their equal-time
counterparts, display non-trivial dynamics \cite{cft3}. In what
follows, we chart out the result for these correlation functions for
the CFT vacuum given by
\begin{eqnarray}
G_n(x_1,x_2) &=& \langle 0|\phi(w_1,\bar w_1) U_n^{\dagger}
\phi(w_2,\bar w_2) U^n |0\rangle \nonumber\\
\label{uetcf1}
\end{eqnarray}
The holomorphic part of this correlator yields after mapping to the
complex plane,
\begin{eqnarray}
G_n^{\rm hol}(x_1,x_2) &=& \prod_{j=1,2} \left( \frac{\partial
w_j}{\partial z_j}\right)^{-h} \left(\frac{\partial z_{n2}}{\partial
z_2}\right)^{h}
\frac{1}{(z_{n2}-z_1)^{2h}} \nonumber\\
&=& \left(\frac{2\pi}{L}\right)^{2h}  \frac{(z_1 z_{2})^h}{\left(
\tilde a_n z_2 +\tilde b_n - \tilde c_n z_1 z_2 - \tilde d_n z_1 \right)^{2h} } \nonumber\\
\label{uetcft2}
\end{eqnarray}
The antiholomorphic part can be computed in a similar manner with
$z_i \to \bar z_i$. Thus one finally gets for $h=\bar h$
\begin{eqnarray}
G_n (x_1,x_2) &=& \left(\frac{2\pi}{L}\right)^{4h} \frac{1}{|
\tilde a_n z_2 +\tilde b_n - \tilde c_n z_1 z_2 - \tilde d_n z_1|^{4h} } \nonumber\\
\label{uetcft3}
\end{eqnarray}
where $z_i= \exp[2\pi i x_i/L]$. Defining the dimensionless center
of mass and relative coordinates as $x_{\rm cm}= \pi(x_1+x_2)/L$ and
$x_{\rm rel}= \pi (x_1-x_2)/L$, analytically continuing to real
time, and substituting Eqs.\ \ref{numexp2}, \ref{numexp3} and
\ref{numexp4} in Eq.\ \ref{uetcft3}, one obtains
\begin{widetext}
\begin{eqnarray}
G_n(x_{\rm cm},x_{\rm rel}) &=& \left(\frac{\pi}{2L}\right)^{4h}
\left[ \left(\alpha \cos(x_{\rm cm}) + \cos(x_{\rm rel}) \right)
\sinh (n\theta) + \sqrt{\alpha^2-1} \cosh(n \theta) \sin(x_{\rm
rel}) \right ]^{-4h} \quad {\rm heating}
\nonumber\\
&=& \left(\frac{\pi}{2L}\right)^{4h} \left[ \left(\alpha \cos(
x_{\rm cm}) + \cos(x_{\rm rel}) \right) \sin (n\theta) +
\sqrt{1-\alpha^2}\cos(n \theta) \sin(x_{\rm rel}) \right ]^{-4h}
\quad {\rm
non-heating}  \label{uetcft4} \\
&=& \left(\frac{\pi}{2L}\right)^{4h} \Big|n s [\cos(x_{\rm cm})+
\cos(x_{\rm rel})]\Big|^{-4h} \, {\rm transition \, line} \nonumber
\end{eqnarray}
\end{widetext}
where we have assumed $h=\bar h$.

\begin{figure}
\rotatebox{0}{\includegraphics*[width= 0.52 \linewidth]{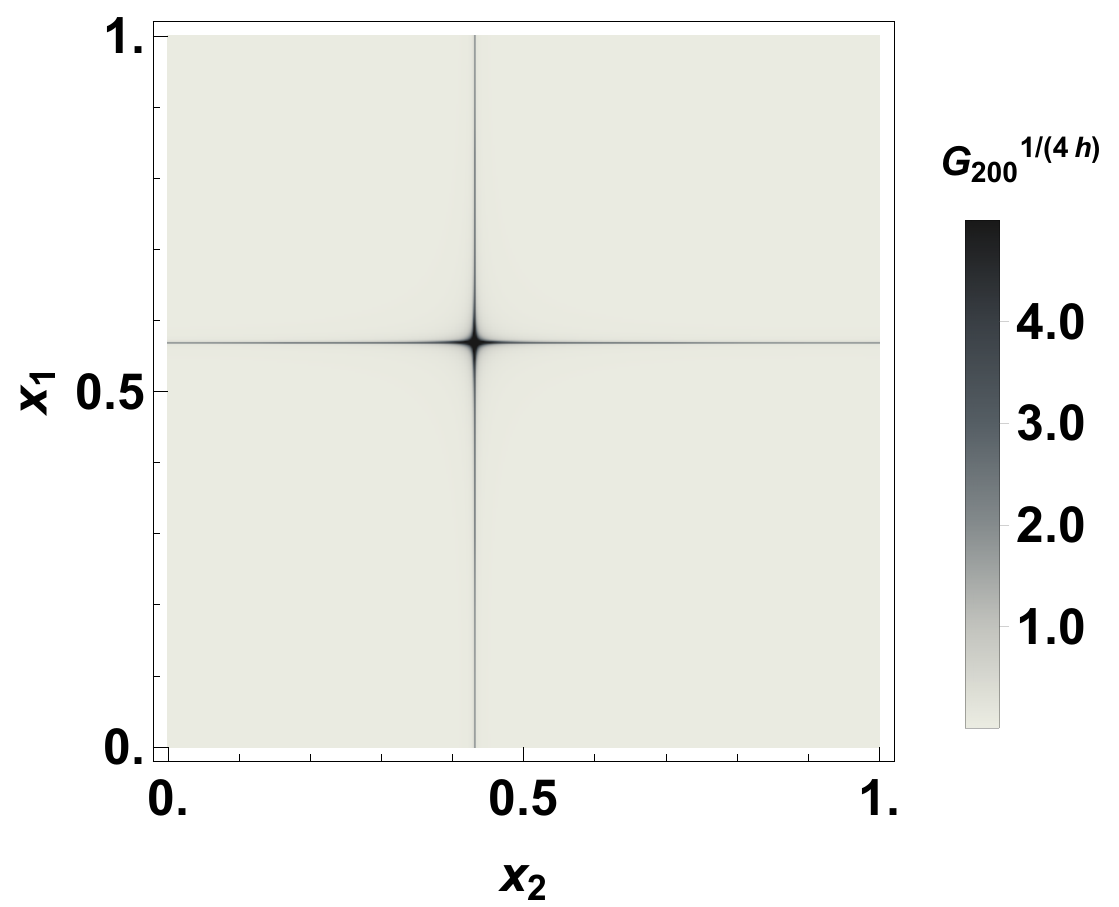}}
\rotatebox{0}{\includegraphics*[width= 0.46 \linewidth]{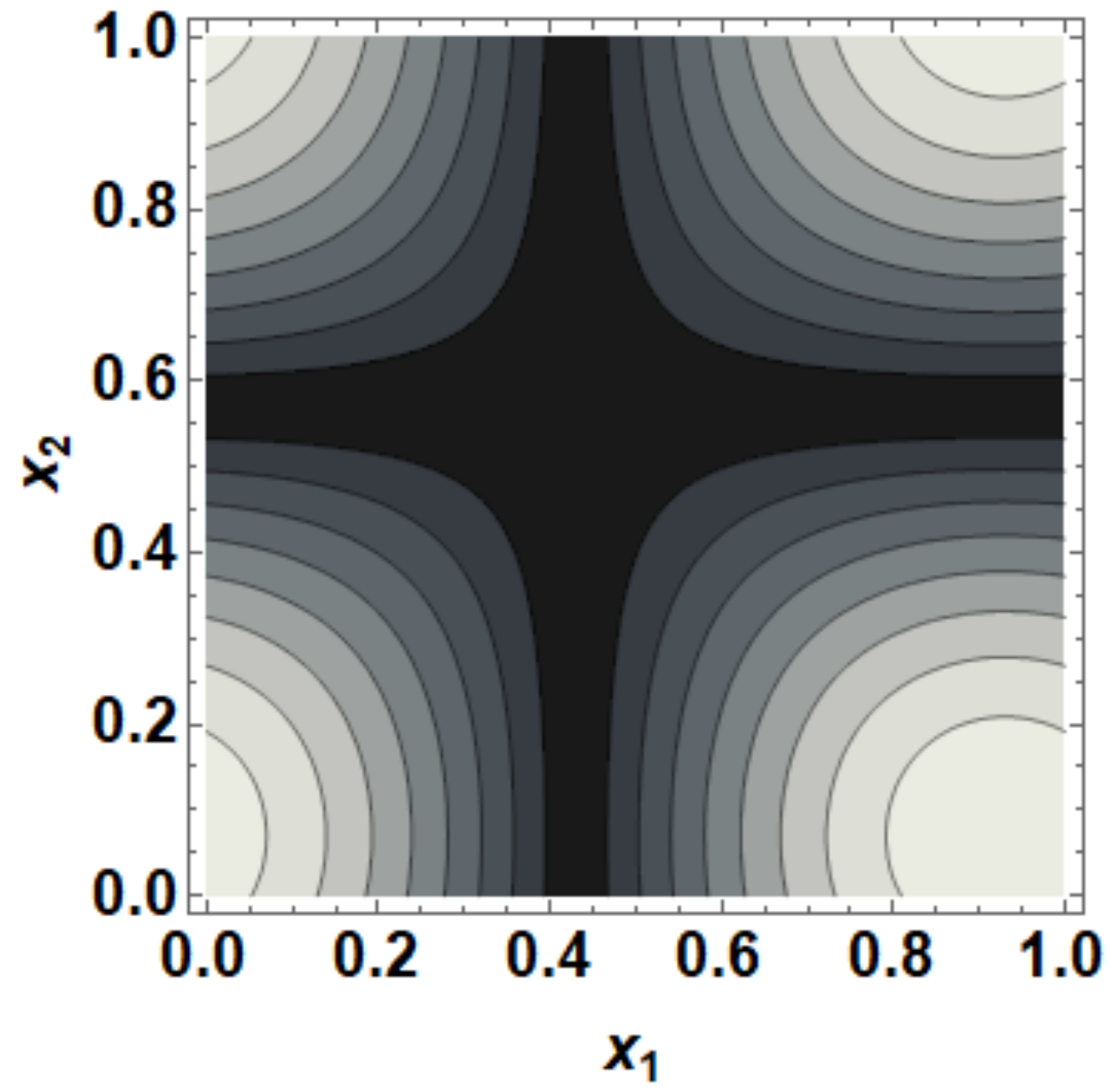}}
\caption{(Color online) Left Panel: Plot of $G$ after $n=10$ drive
cycles for the heating phase($\omega_D=40$, $\delta f=0.98$) as a
function of $x_1/L$ and $x_2/L$ showing the position of the peaks of
$G_n$ in the of $x_1-x_2$ plane. Right panel: Plot of $f(x_1,x_2) =
\alpha \cos x_{\rm cm}+ \sqrt{2} \sin(x_{\rm rel} +\pi/4)$ as a
function of $x_1/L$ and $x_2/L$ for the same parameter values
showing the position of zeros(darker shades) almost coinciding with
the position of peaks in the left panel. For all plots, $f_0=10$.
See text for details.} \label{fig7}
\end{figure}

A straightforward analysis of $G_n(x_{\rm cm}, x_{\rm rel})$ in the
heating phase reveals that it will display peaks, in the large $n$
limit, when
\begin{eqnarray}
\cos(x_{\rm cm}) &=& -\sin(x_{\rm rel}
+\arcsin(1/\alpha))\label{condcorr1}
\end{eqnarray}
For $x_{\rm rel}=0$, we find that the peak position coincide with
those of $E_n(x)$. In general, the position of the peaks trace a
curve in the $(x_1,x_2)$ space and can be tuned by changing
$\alpha$; for $\alpha \to \infty$, the peaks only occurs when
$x_{\rm cm} = x_{\rm rel} + \pi/2$. Thus this phenomenon constitutes
another example of emergence of spatial structure in driven CFTs
which was initially found by analysis of $E_n(x)$ \cite{cft4}. For
all pairs of values $(x_{1}, x_{2})$ which do not satisfy Eq.
\ref{condcorr1}, $G_n$ decays exponentially with $n$ in the large
$n$ limit: $G_n \sim \exp[-4 n h]$. This behavior is clearly seen in
the left panel of Fig.\ \ref{fig7} where $G_n(x_1,x_2)$ is plotted
as a function of $x_1/L$ and $x_2/L$. The position of the divergence
of $G_n(x_1,x_2)$ coincides with the solution of Eq.\
\ref{condcorr1} as can be seen from the right panel of Fig.\
\ref{fig7}.

\begin{figure}
\rotatebox{0}{\includegraphics*[width= 0.49 \linewidth]{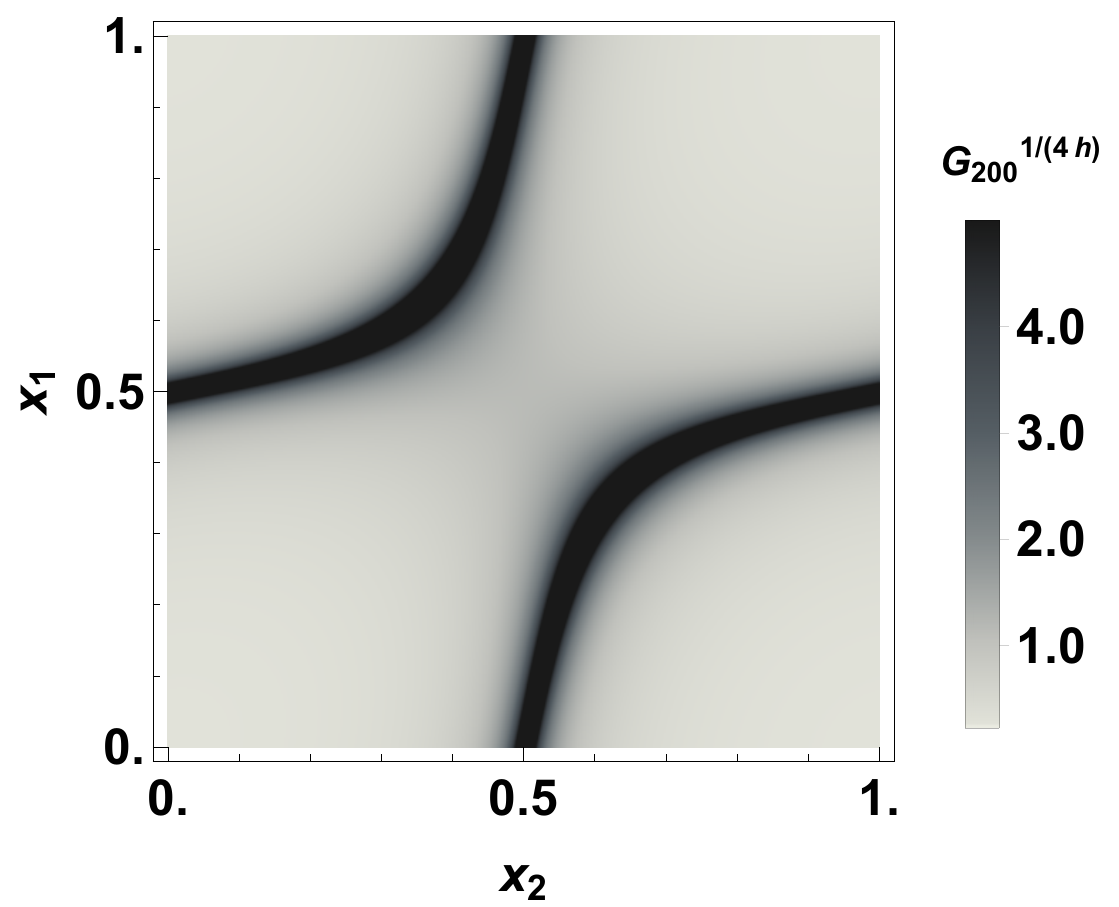}}
\rotatebox{0}{\includegraphics*[width= 0.49 \linewidth]{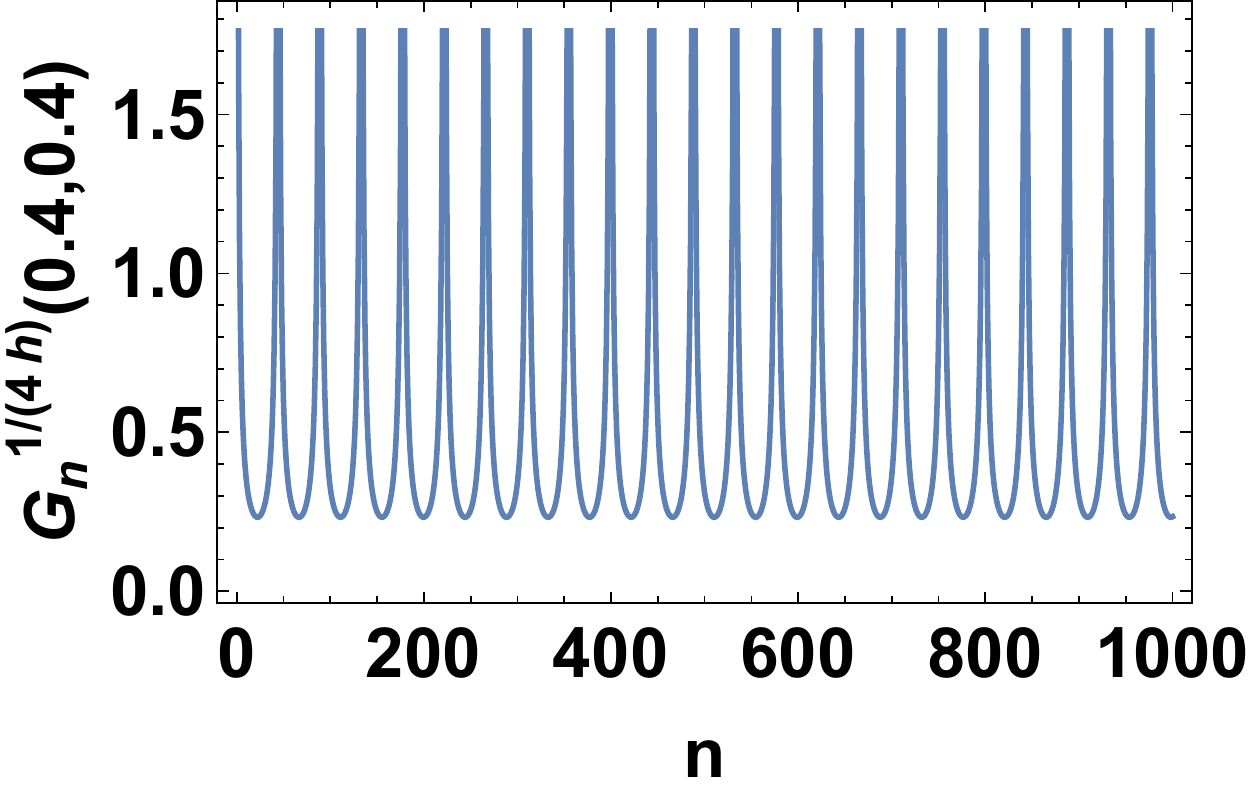}}
\caption{(Color online)Left Panel: Plot of $G_n(x_1,x_2)$ after
$n=10$ drive cycles for the non- heating phase
($\omega_D=100$,$\delta f=1.5$). Right panel: Plot of $G_n(x_1=0.4
L, x_2=0)$ as a function of $n$ for $\omega_D=100$ and $\delta
f=1.5$ showing multiple divergences as a function of $n$ as
predicted by Eq.\ \ref{condcorr2}. For all points $f_0=10$ and
$\pi/L$ is set to unity. See text for details.} \label{fig8}
\end{figure}

\begin{figure}
\rotatebox{0}{\includegraphics*[width= 0.49 \linewidth]{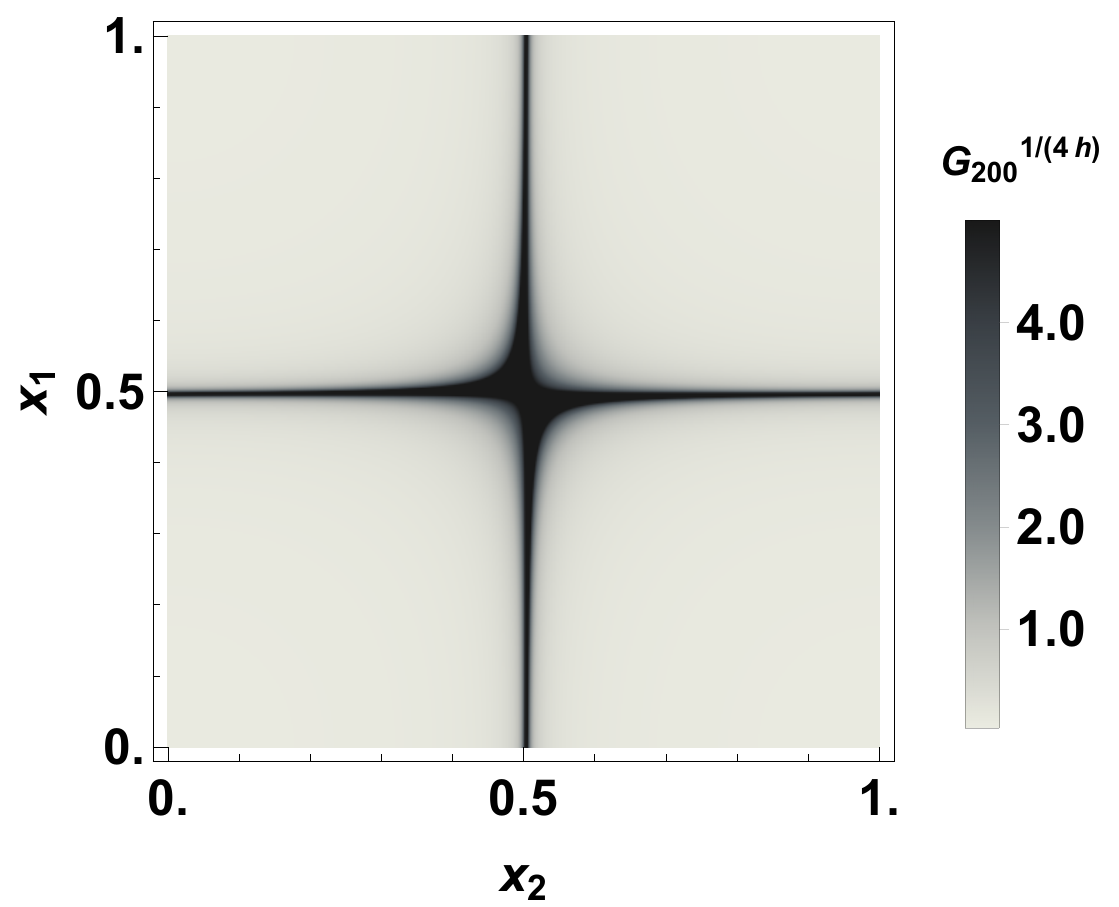}}
\rotatebox{0}{\includegraphics*[width= 0.49 \linewidth]{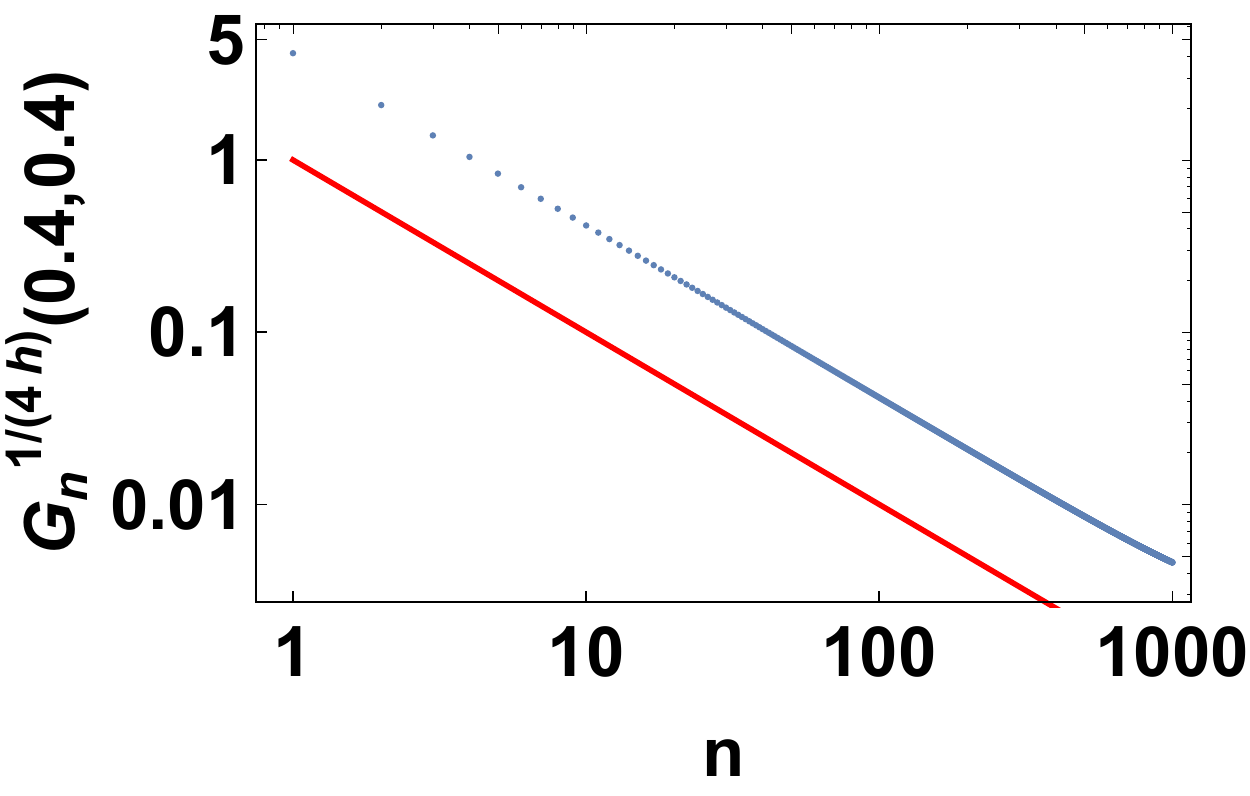}}
\caption{(Color online)Left Panel: Plot of $G_n^{1/4h}(x_1,x_2)$
after $n=10$ drive cycles when the system is on the transition line
($\omega=100$,$\delta f=1$). Right panel: Plot of
$G_n^{1/4h}(x_1=0.99 L, x_2=0)$ as a function of $n$ for
$\omega_D=100$ and $\delta f=1$ showing linear decay with $n$. The
blue dots indicate values of $G_n^{1/4h}$ while the red line is the
linear fit. For all points $f_0=10$ and $\pi/L$ is set to unity. See
text for details.} \label{fig9}
\end{figure}

In contrast, the peaks of $G_n(x_1,x_2)$ in the non-heating phase do
not occur at an fixed positions independent of $n$ in the large $n$
limit. Here the divergences for occur when $n \theta = m \pi$ (for
integer $n$ and $m$) if $x_1=x_2$; for all $x_1 \ne x_2$, they occur
when $n$, $x_1$, and $x_2$ satisfies
\begin{eqnarray}
\frac{\alpha \cos(x_{\rm cm})+ \cos(x_{\rm rel})}{\sqrt{1-\alpha^2}
\sin(x_{\rm rel})} &=& -\cot(n \theta) \label{condcorr2}
\end{eqnarray}
These divergences constitute emergent spatial singularities of the
unequal time correlation function in the non-heating phases and have
no analog in energy density of the system. Such divergences, which
also showed up for equal-time correlation function in the botom left
panel of Fig.\ \ref{fig7}, can be clearly seen in the left panel of
Fig.\ \ref{fig8} where $[G_n(x_1,x_2)]^{1/4h}$ is plotted as a
function of $x_1$ and $x_2$ for $n=30$. The right panel shows
multiple divergences of $G_n^{4(h +\bar h)}$ for $x_1=0.4$ and
$x_2=0$ as a function of $n$. Finally, we note that on the
transition line $G_n(x_1,x_2)$ can diverge if $\cos(x_{\rm
cm})=-\cos(x_{\rm rel})$, {\it i.e.}, for $x_1=L/2$ or $x_2=L/2$. On
the transition line, for large $n$ one finds a $1/n^{4(h+\bar h)}$
decay of the correlator for all $(x_1, x_2)$ except when $x_1=L/2$
or $x_2=L/2$ in which case it diverges. The behavior of
$G_n(x_1,x_2)$ on the transition line as a function of $x_1$ and
$x_2$ is shown in the left panel of Fig.\ \ref{fig9} showing line of
divergences at $x_1=L/2$ and $x_2= L/2$. For $x_1,x_2 \ne L/2$,
$G_n^{1/4h}$ decays linearly with $n$ as shown in the right panel of
Fig.\ \ref{fig9} for $x_1/L=0.99$ and $x_2=0$. These features,
obtained from exact numerics, confirms the analytic prediction of
the first order FPT.

\subsection{Entanglement}

In this subsection, we consider the evolution of the $m^{\rm th}$
Renyi entropy $S_n^m$ after $n$ drive cycles starting from $|h,\bar
h\rangle$ states. We note that the ground state has $S_n^m= S_0^m$
since the state does not change under the drive in the cylindrical
geometry. This entanglement can be computed most simply by
considering the correlation of the twist operators ${\mathcal
T}(w,\bar w)$ \cite{cardyref1}. Here we shall concentrate on the
$m^{\rm th}$ Renyi entropy which is given, after $n$ cycles of the
drive, in terms of the twist operator ${\mathcal T}_m(w, \bar w)$ as
\begin{eqnarray}
S_n^m(\ell) &=& \frac{1}{1-m} \ln \alpha_n^m(\ell), \quad
\alpha_n^m(\ell) = {\rm Tr} \rho_n^m(\ell) \nonumber\\
\alpha_{n}^m(\ell) &=& \frac{\langle h,\bar h|U^{n \dagger}
{\mathcal T}(w_1,\bar w_1) {\mathcal T}(w_2,\bar w_2) U^n |h,\bar
h\rangle}{\langle h,\bar h|h,\bar h\rangle} \label{ee1}
\end{eqnarray}
where $\rho_n(\ell)$ is the reduced density matrix of state after
$n$ cycles of the drive corresponding to an initial state $|h,\bar h
\rangle$, $\ell$ is the spatial dimension of the subsystem, $w_i
(\bar w_i) = +(-) i x_i$, we choose $x_1=0$ and $x_2=\ell$, and the
twist operator ${\mathcal T}_m$ represents a primary field with
dimension $h_m= c(m-1/m)/24$ \cite{cardyref1}. In what follows, we
shall focus on the half-chain entanglement entropy which corresponds
to $\ell = L/2$ for the sake of simplicity; we note however, that
the method can yield results for arbitrary $\ell$.

As before we evaluate the holomorphic part of the
$\alpha_n^m(\ell=L/2)$. This is given by
\begin{eqnarray}
\alpha_n^m &=& \lim_{z_3 \to \infty, z_4 \to 0}\left(\frac{2
\pi}{L} \right)^{2 h_m } \frac{C_{4n}^{m \rm hol}
e^{\pi i h_m } }{\left( \tilde d_n^2  - \tilde c_n^2 \right)^{2h_m } } \nonumber\\
C_{4n}^{m \rm hol} &=& \lim_{z \rightarrow \infty} z^{2h} \langle 0|
\phi(z) {\mathcal T_m}(z_{n1} ) {\mathcal T_m}(z_{n2})
\phi(0)|0\rangle \label{ee3}
\end{eqnarray}
The computation of $C_{4n}^m = C_{4n}^{m \,\rm hol}C_{4n}^{m \, \rm
anti-hol}$ and hence $S_n^m$ thus reduce to a problem similar to
that worked out for the correlation function. However, here the
operator dimension of ${\mathcal T}_m$ are different from $h$, and
hence all the expressions obtained in Sec.\ \ref{corrfns} can not be
directly used. Nevertheless, the computation procedure is similar,
and we present the main results here. We find that $C_{4n}^m$ can
once again be written in terms of sum of contribution over conformal
blocks. In the perturbative limit, where $|1-y_n|, |1-\bar y_n| \ll
1$, one has
\begin{widetext}
\begin{eqnarray}
&& C_{4n}^m =  z_{1n}^{-h -h_m} \bar z_{1n}^{ -\bar h -\bar h_m}
{\mathcal F}'(1-y_n; 1-\bar y_n), \quad {\mathcal F}' = \sum_p C_{h
h p} C_{h_m h_m p} \mathcal{V}_p
\bar {\mathcal V}_{\bar p}\label{cftresen}\\
&& \mathcal{V}_p(y_n,h, h_m) = (1-y_n)^{h_p-h-h_m} \sum_k F_k
(1-y_{n})^k, \quad  \bar {\mathcal V}_{\bar p}(\bar y_n,\bar h, \bar
h_m) = (1-\bar y_n)^{\bar h_p-\bar h -\bar h_m} \sum_k F_k (1-\bar
y_{n})^k,\nonumber\\
&& \sum_k F_k x^k  = 1 + \frac{h_p}{2} x + \frac{h_p \left(h_p
\left(h_p \left(c+8 h_p+8\right)+2 (c+2 h-4)\right)+c-4 h\right)+4
h_m \left(h_p \left(h_p+4 h-1\right)+2 h\right)}{8 h_p \left(c+8
h_p-5\right)+4 c} x^2 + {\mathcal{O}}(x^3) \nonumber
\end{eqnarray}
\end{widetext}
We note that this perturbative result is expected to be accurate for
large $n$ and in the non-heating phases or on the transition line as
discussed earlier. Also in these cases since $|1-y_n|, |1-\bar y_n|
\ll 1$, only the leading term in the sum may be retained.
Substituting Eq.\ \ref{cftresen} in Eq.\ \ref{ee3} and after some
straightforward algebra one obtains, assuming $h=\bar h$ and using
$h_m= \bar h_m$
\begin{widetext}
\begin{eqnarray}
\alpha_n^m (L/2) \simeq \left(\frac{2 \pi}{L}\right)^{4h_m} \sum_p
C_{h h p} C_{h_m h_m p} 4^{\tfrac{h_p+\bar{h}_p}{2} -h_m -h} \left(
\frac{\tilde a_n + \tilde b_n}{c_n + d_n}\right)^{\bar{h}_p-h_p}
|\tilde c_n^2-\tilde d_n^2 |^{2(h-h_m)} (\tilde d_n^2 -
\tilde c_n^2)^{-h_p} (\tilde a_n^2 - \tilde b_n^2)^{-\bar{h}_p} \nonumber\\
\label{entreq1}
\end{eqnarray}
\end{widetext}
If we retain only the identity contribution which is universal, then
using $C_{aaI} = 1$ we can simplify,
\begin{eqnarray}
\alpha_n^m (L/2)_{\rm univ} &\simeq& \left(\frac{2 \pi}{L}\right)^{4h_m} 4^{ -h_m -h} \nonumber\\
&& \times | \tilde c_n^2-\tilde d_n^2 |^{2(h-h_m)} \label{entreq2}
\end{eqnarray}
This universal contribution, after analytic continuation to real
time, yields a simple expression for the evolution of $\delta S_n^m
= S_n^m- S_{n=0}^m$, given by
\begin{eqnarray}
\delta S_n^m(L/2)_{\rm univ} \simeq  \frac{2(h-h_m)} {1-m} \ln
|c_n^2 -d_n^2| \label{identityblocken}
\end{eqnarray}
In the non-heating phase and on the transition line, this yields,
after analytic continuation to real time,
\begin{widetext}
\begin{eqnarray}
\delta S_n^m(L/2)_{\rm univ} & \simeq & \frac{h-h_m}{1-m} \ln \Big
[\left(1+ \alpha^2 \cos 2 n \theta \right)/\sqrt{1-\alpha^2}\Big],
\quad {\rm non-heating}
\nonumber\\
& \simeq & \frac{h-h_m}{1-m} \ln(1+4 n^2 s^2),\quad {\rm transition
\, line} \label{entangres}
\end{eqnarray}
\end{widetext}
Thus the oscillatory behavior of $\delta S_n$ in the non-heating
phase and its logarithmic growth on the transition line are expected
qualitative features that are reproduced in this perturbative
approach. We note that the linear growth of the entanglement in the
hyperbolic phase is also reproduced by this procedure; however, here
we can not ascertain the accuracy of this result since the
perturbation expansion of ${\mathcal F}$ over conformal blocks are
uncontrolled.

In the large central charge limit, for states with fixed conformal
dimensions, we can use the global block approximation to express
(assuming $h_m=\bar h_m$ and $h=\bar h$)
\begin{widetext}
\begin{eqnarray}
&& \alpha_n^m (L/2) \simeq  \left(\frac{2 \pi}{L}\right)^{4 h_m}
\sum_{p\neq {\mathbb I}} C_{h h p} C_{h_m h_m p}
4^{\tfrac{h_p+\bar{h}_p}{2} -h_m -h}
\left( \frac{\tilde a_n + \tilde b_n}{\tilde c_n + \tilde d_n}\right)^{\bar{h}_p-h_p}
(\tilde d_n^2 - \tilde c_n^2)^{-h_p} (\tilde a_n^2 - \tilde b_n^2)^{-\bar{h}_p}\nonumber\\
&& \times |\tilde c_n^2-\tilde d_n^2 |^{2(h-h_m)}\times
{}_2F_1\left( h_p-h+h_m , h_p-h+h_m;2h_p; 1-y_n\right){}_2F_1\left(
\bar h_p-h+h_m , \bar h_p-h+h_m;2\bar h_p; 1-\bar
y_n\right).\label{entreqC1}
\end{eqnarray}
\end{widetext}
Defining $\alpha_n^{m 0}(L/2) = (2\pi/L)^{4h_m} 4^{-(h_m + h)}$ and
$\delta S_n^m = \ln[ \alpha_n^m (L/2)/\alpha_n^{m 0} (L/2)]/(1-m)$,
we find, using Eq.\ \ref{entreqC1},
\begin{widetext}
\begin{eqnarray}
&& \delta S_n^m \simeq  \frac{1}{1-m} \ln \Big[ \sum_{p\neq {\mathbb
I}} C_{h h p} C_{h_m
h_m p} 4^{\tfrac{h_p+\bar{h}_p}{2}} \left( \frac{\tilde a_n + \tilde b_n}
{\tilde c_n + \tilde d_n}\right)^{\bar{h}_p-h_p}
(\tilde d_n^2 - \tilde c_n^2)^{-h_p} (\tilde a_n^2 - \tilde b_n^2)^{-\bar{h}_p}\label{entreqC2} \\
&& \times |\tilde c_n^2- \tilde d_n^2 |^{2(h-h_m)}\times
{}_2F_1\left( h_p-h+h_m , h_p-h+h_m;2h_p; 1-y_n\right){}_2F_1\left(
\bar h_p-h+h_m , \bar h_p-h+h_m;2\bar h_p; 1-\bar y_n\right).\Big]
\nonumber
\end{eqnarray}
\end{widetext}
This provides the drive induced contribution to the half-chain
$m^{\rm th}$ Renyi entropy after $n$ drive cycles.

Finally we consider the case for large $c_0$ CFTs such that $h/c_0
\gg h_m/c_0 $ with $h/c_0$ and $h_m/c_0$ held fixed. Here once
again, it is possible to obtain analytic expression of
$\alpha_n^m(\tfrac{L}{2})$ using the monodromy block Eq.\
\ref{cftlargec}. The universal contribution (from the identity
block) non-perturbative in $n$ is given by,
\begin{widetext}
\begin{eqnarray}
\alpha_n^m(L/2)_{\rm univ} &=& \left( \frac{2\pi a_0}{L}\right)^{4
h_m} |c_n^2-d_n^2|^{-2h_m(a_0+1)} |a_n^2-b_n^2|^{2h_m(a_0-1)}
\Big|\left( \frac{a_n+b_n}{c_n+d_n}\right)^{a_0} -
\left(\frac{a_n-b_n}{c_n-d_n}\right)^{a_0} \Big|^{-4 h_m}
\label{entangcft1} \\
\delta S_n^m (L/2)_{\rm univ} &=& \frac{1}{1-m} \Big(4h_m \ln \Big|
\left(\frac{a_n+b_n}{c_n+d_n}\right)^{a_0} -
\left(\frac{a_n-b_n}{c_n-d_n}\right)^{a_0} \Big| +2 h_m(a_0+1) \ln
|c_n^2-d_n^2| + 2h_m (1-a_0)\ln |a_n^2-b_n^2|. \Big)\nonumber
\end{eqnarray}
\end{widetext}
where we have analytically continued to real time.  We note here
that the exact conformal block contribution can also be determined
numerically using the Zamolodchikov recursion relations \cite{zamu};
however, we are not going to address this in this work.

\section{Relation to lattice models}
\label{secrel}

In this section, we relate our results obtain using conformal field
theory in the last section to those obtained by exact numerics on a
specific lattice model. The model chosen is the sine-square deformed
(SSD) fermionic model whose Hamiltonian is given by
\cite{ssdpapers,cft1}
\begin{eqnarray}
H_{\rm SSD} &=& [H_0 + (H_+ + H_-)/2]/2  \nonumber\\
H_0 &=& -\sum_j J(c_j^{\dagger} c_{j+1} +{\rm h.c}) \label{latticeham} \\
H_{\pm} &=& -\sum_j {J_1} e^{ij \delta} (c_j^{\dagger}
c_{j+1} +{\rm h.c.}) \nonumber
\end{eqnarray}
where $c_j$ is the fermion annihilation operator on site $j$,
$\delta=2 \pi/L$, $L$ is the chain length, the lattice spacing is
set to unity, we have assumed that the system to be at half-filling,
and $J$ is the hopping strength of the fermions. In what follows, we
shall use periodic boundary condition for this Hamiltonian. We note
that due to the local phase factor $\exp[\pm i \delta j]$, $H_{\pm}$
are not Hermitian operators. However, their sum is still Hermitian
and leads to
\begin{eqnarray}
H_{\rm SSD} &=& - \sum_j J \Lambda_j (c_j^{\dagger} c_{j+1}+
+{\rm h.c.}) =\sum_j h_j \nonumber\\
\Lambda_j &=& 1+ 2 J_1\cos\left(
\frac{2\pi}{L}(j-1/2)\right)/J\label{ssdham2}
\end{eqnarray}
where $J$ is the hopping strength of the fermions. In what follows,
we shall implement the drive via a time-dependent hopping $J \to
J(t)= J_1+J_0\cos(\omega_D t)) + \delta J $. It is well known
\cite{ssdpapers,cft1} that the low-energy sector of Hamiltonian can
be expressed as
\begin{eqnarray}
H &=& \frac{2 \pi}{L} (J(t) L_0 + J_1(L_1+L_{-1})/2)
\nonumber\\
&& +{\rm anti-holomorphic \, part}  \label{lattice2}
\end{eqnarray}
so that one can identify $\delta f = \delta J +J_1$ and $J_0=f_0$.
In what follows, we shall scale all energies by $J_1$ and compute
the time evolution of the instantaneous energy density of the
system, $E_n(x)$ after $n$ cycles of the drive as follows.

To this end, we first compute $U(T,0) = T_t \exp[- i \int_0^T dt
H_{\rm SSD}(t)/\hbar]$ numerically. The procedure for this identical
to the one carried out in Sec.\ \ref{secdrive} and involves
decomposition of $U$ into $N$ time steps of width $\delta t= T/N$:
$U(T,0) = \prod_{j=0..N-1} U_j$, where $U_j= U(t_j+\delta t,t_j)$.
The width $\delta t$ is chosen such that $H_{\rm SSD}$ does not vary
appreciably within this interval. One then diagonalizes $H_j$ and
expresses $U_j$ in terms of its eigenvalues and eigenvectors. The
matrix $U$ is then constructed by taking product over all $U_j$s.
Finally one diagonalizes $U$ to obtain the Floquet eigenvalues
$\epsilon_m^{F \,\rm SSD}$ and eigenvectors $|m_{\rm SSD}\rangle$.
In terms of these after n drive cycles, the wavefunctions of the
driven chain can be written as
\begin{eqnarray}
|\psi_n\rangle &=& \sum_m e^{-i \epsilon_m^{F \, \rm SSD} n
T/\hbar}c_{m} |m_{\rm SSD}\rangle \label{wavssd}
\end{eqnarray}
where $c_m =\langle m_{\rm SSD}|\psi_{\rm init}\rangle$ denotes the
overlap of the Floquet eigenstates with the initial state. The
initial state is chosen to be one of the primary CFT states; for the
lattice model studied here, these states are tabulated in Ref.\
\onlinecite{xxcft}. We note since $f_0 \gg 1$, the initial primary
states corresponding to $H_{\rm SSD}$ is expected to be accurately
described by those charted in Ref.\ \onlinecite{xxcft}. Here we
choose the state corresponding to $h=\bar h=1/2$ which for the
lattice correspond to the state
\begin{eqnarray}
|\psi_{\rm init}\rangle &=& c_{k_F + \pi/L}^{\dagger}
c_{-k_F-\pi/L}^{\dagger} |{\rm FS}\rangle \label{init1}
\end{eqnarray}
where $|{\rm FS}\rangle$ is the half-filled Fermi sea. Thus
$|\psi_{\rm init}\rangle$ corresponds to two particles populating
the lowest available energy states over the half-filled Fermi sea
\cite{xxcft}. One then computes the instantaneous energy density at
any given site as $E_n(j)= \langle \psi_{\rm
init}|h_{j}(n)|\psi_{\rm init}\rangle$, where $h_{j}(n)=
U^{\dagger}(nT,0) h_j(t=0)U(nT,0)$. In what follows we shall study
the behavior of $E_n(j)$ and relate it to the corresponding CFT
results obtained in Sec.\ \ref{rpensec}.

\begin{figure}
\rotatebox{0}{\includegraphics*[width= 0.49 \linewidth]{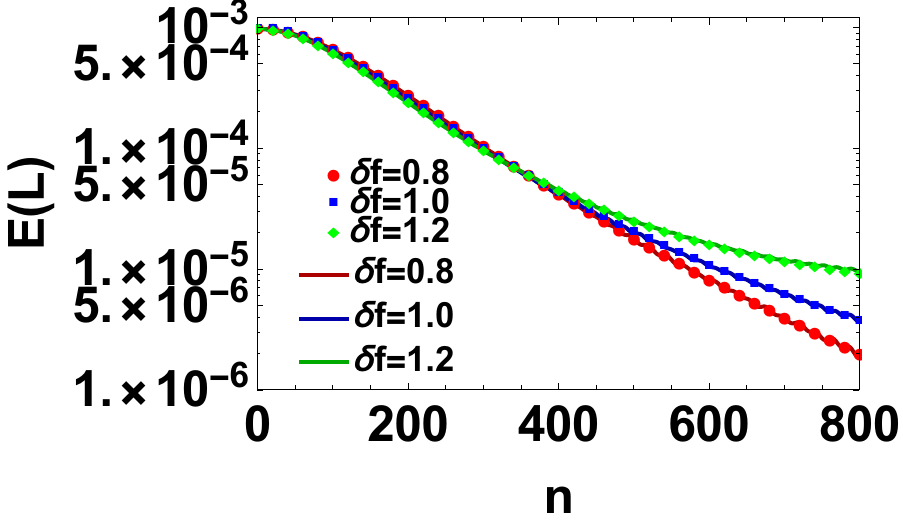}}
\rotatebox{0}{\includegraphics*[width= 0.49 \linewidth]{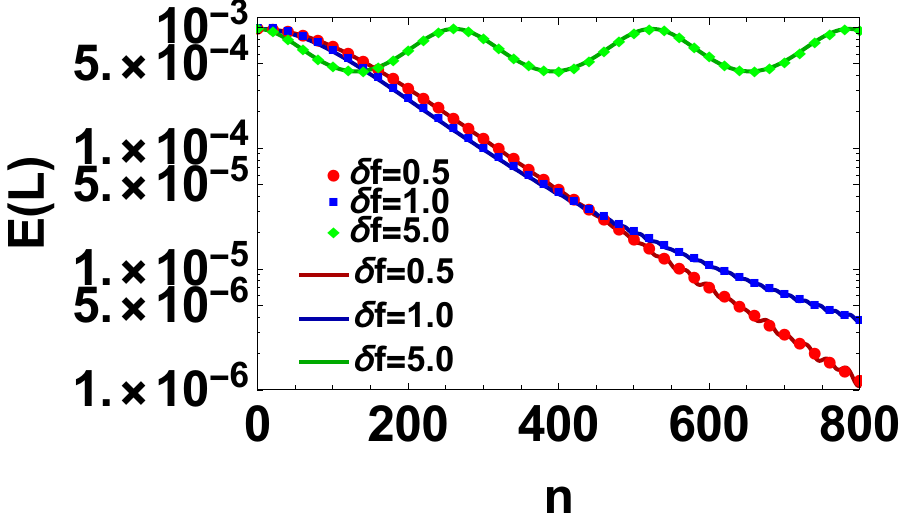}}
\rotatebox{0}{\includegraphics*[width= 0.49 \linewidth]{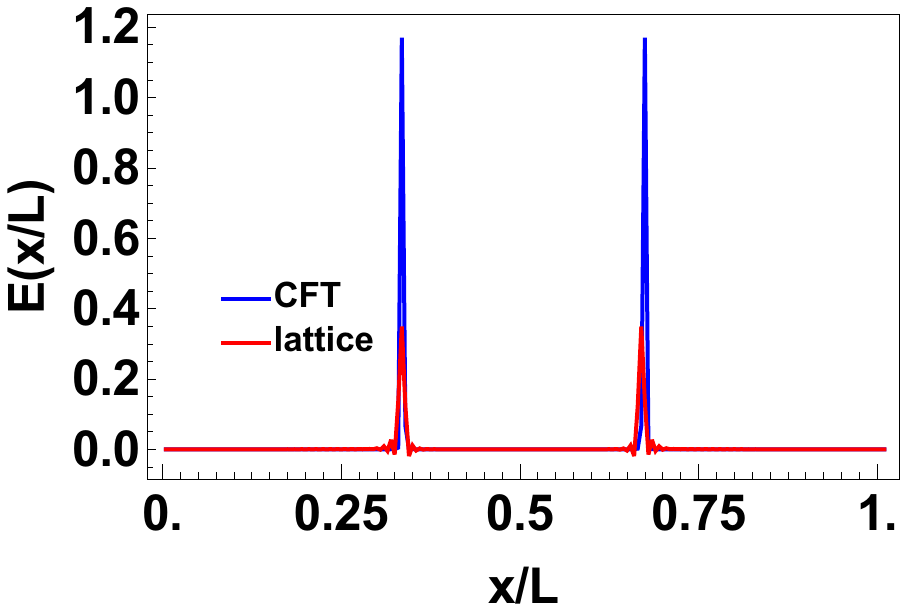}}
\rotatebox{0}{\includegraphics*[width= 0.49 \linewidth]{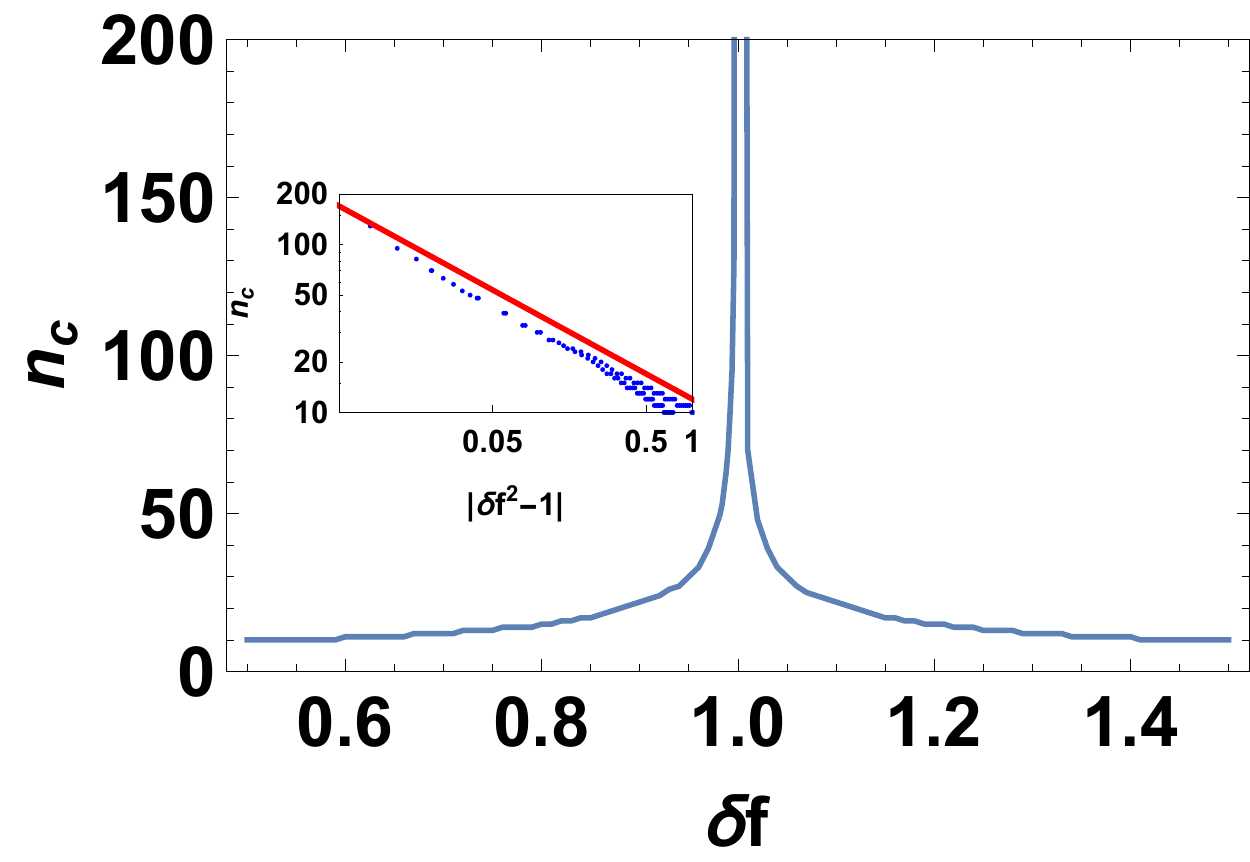}}
\caption{(Color online)Top Left Panel: Plot of $E_n(L)=E_n(0)$ as a
function of $n$ as obtained from lattice (solid lines) and CFT
(dotted lines) calculations showing universal behavior for $n \le
n_c \simeq 200$ for several representative values of $\delta f$ (in
units of $\pi/L$). Top Right panel: Plot of $E(L)$ as a function of
$n$ obtained from lattice (solid lines) and CFT (dotted lines)
calculations showing deviation oscillatory and decaying behaviors in
non-heating and heating phases and on the transition line. Bottom
left panel: Plot of $E(x)$ as a function of $x$ after $n=1500$ drive
cycles and $\delta f= 5 \pi/L \simeq 0.078$ showing the emergent
spatial structure. Bottom Right panel: Plot of $n_c$ as a function
of the $\delta f$ (in units of $\pi/L$) showing the divergence at
the transition line. The inset shows plots of $\delta n_c$ vs
$\delta f$ on log scale. The red line corresponds to a plot of
$1/\sqrt{|\delta f^2 -1|}$. For all points $\omega_D=40J_1/\hbar$
for which $\delta f_c \simeq \pi/L$ and we have chosen $L=202$. See
text for details.} \label{fig11}
\end{figure}

The results obtained from this procedure is shown in Fig.\
\ref{fig11} for a chain of length $L=200$. The top left panel shows
a plot of $E(L)$ as a function of $n$ for $\omega_D =40 J_1/\hbar$
for several representative values of $\delta f$. At this frequency,
the transition line is at $\delta f_c \simeq 1$. One therefore finds
that below a threshold number of drive cycles, $n_c$, $E(L)$ as
obtained from the lattice shows universal behavior for both the
phases and mimics the behavior of the system on the transition line.
Moreover, these results show an excellent match with the
corresponding CFT results outlined in Eqs.\ \ref{heatingen},
\ref{nonheatingen}, and \ref{tranen} (with $L$ in Eq.\ \ref{hamcft}
identified to the chain length of the lattice Hamiltonian). The long
time behavior of $E(L)$ as a function of $n$ in all the phases is
shown in the top right panel of Fig.\ \ref{fig11}. We find that the
lattice model reflects a clear distinction, as predicted by CFT,
between the behavior of $E(L)$ in the heating and non-heating phases
at long times (large $n$). The bottom left panel shows the energy
density $E(x)$ as a function of $x/L$ in the heating phase
corresponding to $\omega_D=40 J_1/\hbar$ and $ \delta f= 5\pi/L$
after $n=1500$ cycles of the drive. We find that the peak positions
predicted by CFT are correctly captured by the lattice model;
however, the peak amplitudes are lower which is due to lattice
effects that are expected to cause deviation of lattice dynamics
from the CFT results in the heating phase at large $n$ \cite{cft2}.
Finally, in the bottom right panel of Fig.\ \ref{fig11}, we find
that $n_c$ diverges at the transition; this divergence seems to
coincide with the predicted $1/\sqrt{|1-\alpha^2|}$ behavior since
$\alpha \sim 1/\delta f$ for $\omega_D \pi/L \gg 1$.

\section{Discussion}
\label{conc}

In this work, we have studied the dynamics of driven CFTs using a
continuous protocol. Our analysis shows that such a drive protocol,
characterized by amplitudes $f_0, \delta f$ and frequency
$\omega_D$ yield heating and non-heating phases separated by
transition lines. Such phases were obtained for discrete protocols
earlier in Refs.\ \onlinecite{prapaper1} and \onlinecite{cft1} where
the evolution operator $U(T,0)$ admits an exact analytic solution.
In contrast, for continuous drive protocol, there is no exact
analytic result for $U$. We therefore first present the phase
diagram in the limit of large drive amplitude as a function of
$\delta f$ and $\omega_D$ showing several re-entrant transitions
between the heating and the non-heating phases. We also develop an
analytic, albeit perturbative, approach to these driven systems
using FPT. We find that for $\hbar \omega_D \ge \delta f,1$ (in
units of $\pi/L$), the first order FPT results provide an excellent
match with the numerical results. Our analysis allows us to identify
a parameter $\alpha$ as a function of $f_0, \delta f$, $\omega_D$ whose
values determine the phase of the system; for $|\alpha| < (>)1$,
the system is in the non-heating (heating) phase. The transition
lines correspond to $\alpha =\pm 1$.

We have also investigated the return probability, energy density, correlation
function and the Renyi entropies of the driven CFT and have provided
perturbative analytic expressions for several quantities using FPT. 
These expressions provide reasonable match with exact numerics. 
We show that the return probability $P_n$ of a primary state displays decaying
(oscillatory) behavior in the heating (non-heating) phase as
expected. On the transition line $P_n$ shows a power-law decay with
$n$. Our analysis identifies a crossover stroboscopic timescale
$n_c$; for $n \le n_c$, the return probability shows a universal
behavior analogous to that of a system on the transition line. We
show that $n_c \sim 1/\sqrt{|1-\alpha^2|}$ and can thus be tuned by
changing both $\delta$ and $\omega_D$.

For energy density of a primary state, we find emergence of spatial
structure as identified earlier in Ref.\ \onlinecite{cft3}. The
peaks of the energy density for large $n$ occurs at $L/4$ and $3L/4$
when the system is deep inside the heating phase ($|\alpha| \gg 1$);
they move towards $L/2$ as one approaches the transition line
($\alpha =\pm 1$). In contrast, there are no such peaks in the
non-heating phase which are independent of $n$ in the large $n$
limit; here $E_n(x)$ shows an oscillatory behavior as a function of
$x$ for all $n$. For small $n$, we find that $E_n(x)$ obeys
universal behavior similar to that when the system is on the
transition line and identify a crossover time scale till which this
behavior persists. However this phenomenon does not occur if $x$
corresponds to the position of the peaks in the heating phase or on
the transition line.

We have also computed the equal-time correlation functions of
primary fields, $C_n(x_1,x_2)$ of the driven CFT starting from a
primary state. These correlation function requires evaluation of the
four-point function of the CFT and are therefore expressed in terms
of ${\mathcal F}$ which admits decomposition into Virasoro {conformal} 
blocks $\mathcal{V}_p$. The analytical expression of
${\mathcal V}_p$ for arbitrary $(h,h_p)$ does not exist. Here we
have identified several limiting case where analytical results may
be presented. The first of these is the case when the cross ratios
that appear in the argument of ${\mathcal F}$ ($|1-y_n|$ and
$|1-\bar y_n|$ in our case) are small. This limit is applicable for
large $n$ if $|x_1-x_2| \ll L$ for all phases; it is also applicable
for all $x_1$ and $x_2$ in the non-heating phase and on the
transition line provided $n$ is large. Our analysis in this limits
shows emergent peaks in the hyperbolic phase analogous to the energy
density. In addition, we also find emergent spatial structure
indicating a line of divergence in the non-heating phase and on the
transition line. We also provide analytic expression using large
$c_0$ limit for the block with fixed $h, \bar h$ where the Virasoro
blocks can be replaced by the global conformal block. Finally, we
note that using monodromy methods, it is possible to find analytic
expressions of the correlation functions in the large $c_0$  when
the dimension of the primary state $H \gg h$ with $H/c_0$ and
$h/c_0$ held fixed. We point out that this regime may be relevant
for large $c_0$ CFTs used in AdS/CFT correspondence.

The structure of the unequal time correlator $G_n(x_1,x_2)$ in the
presence of the drive provides another example of emergent spatial
structure. We note that here one can study the dynamics starting
from the cylinder vacuum state since the unequal-time correlation
function for such an initial state, in contrast to its equal-time
counterpart, shows non-trivial evolution. Our analysis shows that
$G_n$, in the heating phase, diverges along a curve in the $x_1,x_2$
plane; we provide an analytic expression for this curve within first
order FPT which shows reasonable match with exact numerics. We also
find that in contrast to $E_n(x)$, $G_n$ also shows divergences
along a curve in the non-heating phase; the shape of this curve
depends on $n$ through Eq.\ \ref{condcorr2}. This constitutes an
example of emergent spatial structure in the non-heating phase which
does not exists for $E_n(x)$.

Next, we provide a computation of the half-chain entanglement
entropy ($m^{\rm th}$ Renyi entropy) for the driven CFT staring from
a primary state. The computation of $S_n^m$ is similar to that of
the equal-time correlation function since it can indeed be viewed as
equal time correlation function of the twist operator ${\mathcal
T}_m$ with conformal dimension $h_m$. We express $S_n^m$ in terms of
the conformal blocks and discuss limits in which their analytic
expressions are available. Such a limit constitutes the case of
long-time ($n \gg 1$) limit of $S_n^m$ in the non-heating phase and
on the transition line. Here we show that universal contribution to
${\mathcal F}$ (and hence $S_n^m$) from the identity block has
oscillatory dependence of $n$ in the non-heating phase and a
logarithmic growth on the transition line. We also provide analytic
expression for $S_n^m$ in the large $c_0$ limit where the Virasoro
blocks can be replaced by global conformal blocks. Finally for $h
\gg h_m$, and $c_0 \gg 1$ with $h/c_0$ and $h_m/c_0$ held fixed, we
find analytic expression for $S_n^m(L/2)$ using monodromy methods;
our results here may be relevant to CFTs used in AdS/CFT
correspondence.

Finally, we point out that our results show excellent match with
exact numerics of a 1D lattice model of fermions on a finite chain
of length $L$ with Hamiltonian $H_{\rm SSD}$ (Eq.\ \ref{ssdham2}).
In particular, we find that the emergent spatial structure of the
energy density in the heating phase at long times and its universal
behavior below a crossover scale $n_c$ is accurately reflected in
such lattice dynamics. We note that we study the system in the
presence of a global drive. In a typical lattice system which obeys
Galilean invariance, such a drive does not usually lead to emergent
spatial structure of correlation functions or energy densities. The
fact that we find such an emergent structure here clearly shows the
necessity of a CFT based interpretation of such a dynamics where
space and time are intertwined \cite{cft2}.

In conclusion we have studied driven CFTs using a continuous
periodic protocol and have provided a phase diagram showing
re-entrant transitions between heating and non-heating phases. We have
also studied the return probability, energy density, correlation
functions and Renyi entropies of such a driven CFT starting from
primary states. Our results indicate several features of these
quantities such as the universal behavior of the return probability
and the energy density below a crossover stroboscopic timescale and
emergence of spatial structure in both heating and non-heating
phases as found in the correlation function of primary fields. We
discuss relations of these results to a recently studied lattice
model and find excellent match between exact numerical lattice model
based results with analytic prediction of the CFT.

\section{Acknowledgement}

The authors thank Shouvik Datta and especially, Koushik Ray for several 
stimulating discussions. RG acknowledges CSIR SPM fellowship for support. DD acknowledges  
supports provided by {\small{SERB, MATRICS}} and Max Planck Partner Group grant, 
{\small{MAXPLA/PHY/2018577}}.

\appendix

\section{Floquet perturbation theory}
\label{appa}

In this appendix, we provide details of the Floquet perturbation
theory used in the main text. We begin our analysis starting from
Eq.\ \ref{hamcft} of the main text with $f(t)$ given by Eq.\
\ref{protocol1}. We shall put $\pi/L = \hbar=1$ in this section.

In the limit of large $f_0$, $U_0$, the zeroth order term in the
perturbative expansion of $U$ is given by ,
\begin{eqnarray}
U_0(t,0)&=& e^{-i \int_0^t H_0(t^{\prime}) dt^{\prime}}
\nonumber\\
&=& e^{-i\sigma_z (\frac{f_0}{\omega_D}\sin(\omega_D t)+\delta f t)}
\label{u0eqapp}
\end{eqnarray}
Since $\sin (\omega_D T)=0$, where $T=\frac{2 \pi}{\omega_D}$ is the
time period of the drive, we find (restoring $\pi/L$ and $\hbar$)
\begin{eqnarray}
U_0(T,0)=e^{-i \delta f T \sigma_z},\quad H_F = s \sigma_z /T
\label{u0eq}
\end{eqnarray}
where $s =\arccos(\cos(\delta f T ))$ (where $\pi/L$ is set to
unity) is defined in the main text. We note that $H_F^{(0)}$
reflects the periodicity of $U$.

The first order term for $U$ in the perturbation expansion is given
by
\begin{equation}
U'_1(T,0)=-i \int_0^T dt U_0^{\dagger} H_1 U_0
\end{equation}
Since $H_1 \sim i \sigma_y$ and $U_0$ only depends on $\sigma_z$
(Eq.\ \ref{u0eq}), a straightforward calculation yields
\begin{eqnarray}
U'_1(T,0) &=& =-i(I_1 \sigma_+ + I_1^{\ast} \sigma_-) \nonumber\\
I_1 &=&\sum_m J_m\left(\frac{2 f_0}{\omega_D}\right)\int_0^T dt e^{i (m \omega_D+ 2 \delta f)t}\nonumber\\
&=&\sum_m J_m \left(\frac{2 f_0}{\omega_D }\right) \frac{e^{i s} T
\sin s}{m \pi+ \delta f T}
\end{eqnarray}
where $J_m(x)$ are $m^{th}$ Bessel functions. This leads to Eq.\
\ref{u1eq} and then, following the unitarization procedure discussed
in the main text, to Eq.\ \ref{fpt1} for $H_F^{(1)}$.

\begin{figure}
\rotatebox{0}{\includegraphics*[width= 0.98 \linewidth]{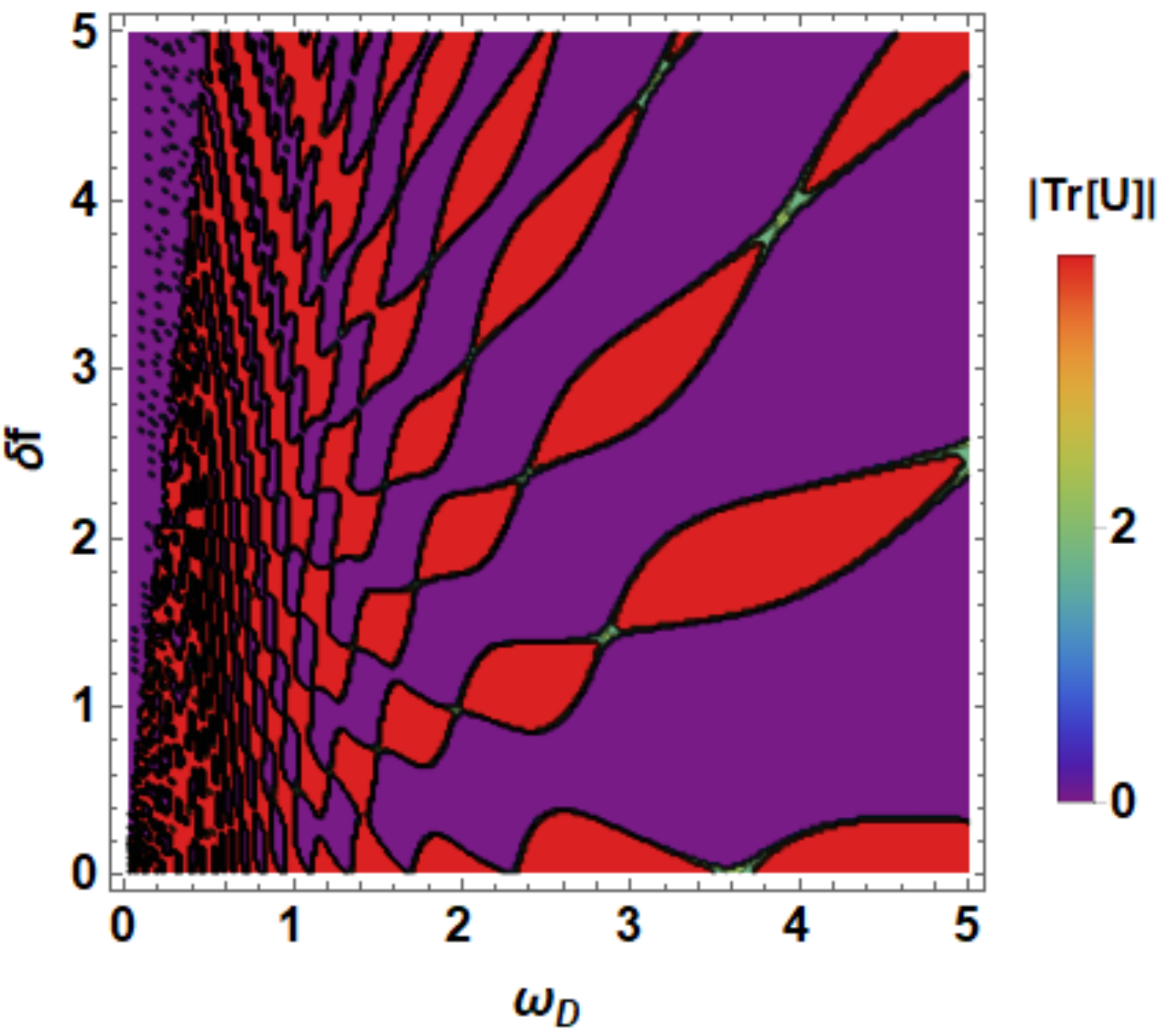}}
\caption{(Color online) Left Panel: Plot of the phase diagram
showing $|{\rm Tr}U(T,0)|$ as a function of the amplitude $\delta f$
and frequency $\omega_D$ as obtained from second order FPT.}
\label{fig12}
\end{figure}

Next, we compute  the second order term in the perturbative
expansion of $U$
\begin{eqnarray}
U'_2(T,0)&=&(-i)^2\int_0^T dt_1 \int_0^{t_1} dt_2 U_0^\dagger(t_1,0)H_1U_0(t_1,0)\nonumber\\
&& \times U_0^{\dagger}(t_2,0) H_1
U_0(t_2,0)\label{usencond}
\end{eqnarray}
Once again using the Pauli matrix dependence of $H_1$ and $U_0$ we
find
\begin{eqnarray}
U'_2 &=&-(\sigma_+ \sigma_-I_2 +\sigma_-\sigma_+ I_2^*)
\label{secondorder} \\
I_2 &=& \sum_{m,n}J_m(x) J_n(x)\int_0^T dt_1 e^{i (m \omega_D
+\delta
f)t_1} \nonumber \\
&& \times \int_0^{t_1} dt_2e^{i (n \omega_D +\delta f)t_2} \nonumber\\
&=& \sum_{m,n}J_m(x) J_n(x)\frac{i T^2}{2 (\pi m + \delta f
T)}[-\delta_{mn}+ \frac{e^{is} \sin s}{\pi n + \delta f T}]
\nonumber
\end{eqnarray}
where $x=2 f_0/\omega_D$. The real and imaginary parts of $I_2$ can
be read off as
\begin{eqnarray}
{\rm Re} [I_2] &=&\sum_{m,n}J_m(x) J_n(x)\frac{T^2}{2(\pi m+ \delta f T)(\pi n+ \delta f T)}\sin^2 s\nonumber \\
{\rm Im} [I_2]&=& \beta = \sum_{m,n} J_m(x)J_n(x)\frac{T^2}{2(\pi m+ \delta f T)}\nonumber\\
&& \times [\delta_{m n} - \frac{\cos s \sin s}{(\pi n + \delta f
T)}] \label{realimag}
\end{eqnarray}
From Eq.\ \ref{realimag} we note that $U_2 - U_1^2/2$ depends only
on ${\rm Im}[I_2]$. Thus to  second order in perturbation theory,
$U_2$ is given by
\begin{eqnarray}
U_2 &=& U_0(I+U'_1+(U'_2-U_1^{'2}/2))=e^{-i H_F^{(2)}T} \nonumber\\
&=& \left(\begin{array} {cc}
e^{-i s}(1+i \beta) & -i \alpha \sin y \\
i \alpha  \sin y & e^{i y}(1-i \beta)
\end{array} \right)
\end{eqnarray}
where $\alpha$ is defined in Eq.\ \ref{alphaeq} in the main text. We
unitarize $U_2$ following the same procedure as
\begin{eqnarray}
H_F^{(2)} &=& \theta^{(2)} \left(n_z^{(2)} \sigma_z + i n_y^{(2)}
\sigma_y \right)/T \nonumber\\
\sin (\theta^{(2)} T) &=& \sqrt{ (\sin s - \beta \cos s)^2 -
\alpha^2 \sin^2 s} \nonumber\\
n_z^{(2)} \sin (\theta^{(2)} T) &=& \sin s -\beta \cos s \nonumber\\
n_y^{(2)} \sin (\theta^{(2)} T) &=& \alpha \sin s \label{secondhf}
\end{eqnarray}
The phase diagram is obtained from Eq.\ \ref{secondhf} by imposing
condition on ${\rm Tr}[\exp[-i H_F^{(2)} T]]$ as discussed in the
main text. This translates to the conditions $|\cos(\theta^{(2)})|>
(<)2$ for the heating (non-heating) phases and
$|\cos(\theta^{(2)})|=2$ on the transition line. The phase diagram,
shown in Fig.\ \ref{fig12} as a function of $\delta f$ and
$\omega_D$ turns out to be qualitatively similar to that obtained
using first order FPT (Fig.\ \ref{fig1} in the main text).

\section{Mobius transformation}
\label{appb}

In this section, we discuss several aspects of the Mobius
transformation corresponding to continuous protocol discussed in
this work. To this end, we first relate to the Mobius transformation
used in Ref.\ \onlinecite{cft1}. It was shown that for an
Hamiltonian $H=H_0 + \tanh(\theta) (H_++H_-)/2$, the transformation
is given by
\begin{equation}
z_{new}=\frac{[(1-\lambda) \cosh(2 \theta)
-(\lambda+1)]z+(\lambda-1) \sinh(2\theta)}{(1-\lambda)
\sinh(2\theta)z+[(\lambda-1)\cosh (2 \theta)-(\lambda+1)]}
\end{equation}
where {$\lambda=\exp[\delta \tau/\cosh(2 \theta)]$} and
$\tau$ is the imaginary time. We show below that this relation is
reproduced for us at every Trotter steps.

To this end, we note that the instantaneous Hamiltonian at any time
$\tau$ is of the form,
\begin{equation}
H=a H_0-\frac{b}{2}(H_++H_-)=a[H_0-\frac{b}{2a}(H_++H_-)]
\end{equation}
where $a$ and $b$ depends on $\tau$. This indicates that for the
non-heating phase one can write $\tanh 2 \theta=b/a$ and {$\lambda=\exp[a \delta \tau/\cosh(2
\theta)]=\exp[\sqrt{a^2-b^2} \delta \tau]$}, since {
$\cosh 2 \theta=a/\sqrt{a^2-b^2}$ and $\sinh 2
\theta=b/\sqrt{a^2-b^2}$}. The Mobius transformation corresponding
to such a Hamiltonian is given by
\begin{widetext}
\begin{equation}
M_1=
\begin{pmatrix}
a_1&b_1\\
c_1&d_1
\end{pmatrix}=\begin{pmatrix}
s_2-s_1 & \frac{b}{\sqrt{a^2-b^2}}(\exp[\sqrt{a^2-b^2}\delta
\tau]-1)\\
- \frac{b}{\sqrt{a^2-b^2}}(\exp[\sqrt{a^2-b^2}\delta \tau]-1) & s_2+
s_1
\end{pmatrix}
\end{equation}
\end{widetext}
where $s_1= (1-\exp[\sqrt{a^2-b^2}\delta \tau])a/\sqrt{a^2-b^2}$ and
$s_2= (\exp[\sqrt{a^2-b^2}\delta \tau]+1)$. Thus we seek a matrix of
the form $M=\frac{1}{2}(a \sigma_z-i b \sigma_y)$ whose exponential
gives the Mobius matrix $M_1$. Since
\begin{widetext}
\begin{equation}
M_2= e^{-\Delta \tau M} = \left(
\begin{array}{cc}
 \cosh \left(\frac{1}{2} \sqrt{a^2-b^2} \delta \tau \right)-\frac{a \sinh \left(\frac{1}{2} \sqrt{a^2-b^2} \delta \tau \right)}{\sqrt{(a-b) (a+b)}} & \frac{b \sinh \left(\frac{1}{2} \sqrt{a^2-b^2} \delta \tau \right)}{\sqrt{(a-b) (a+b)}} \\
 -\frac{b \sinh \left(\frac{1}{2} \sqrt{a^2-b^2} \delta \tau \right)}{\sqrt{(a-b) (a+b)}} & \cosh \left(\frac{1}{2} \sqrt{a^2-b^2} \delta \tau \right)+\frac{a \sinh \left(\frac{1}{2} \sqrt{a^2-b^2} \delta \tau \right)}{\sqrt{(a-b) (a+b)}} \\
\end{array}
\right)
\end{equation}
\end{widetext}
we find $M_2(1,1) = a_1 \exp[\frac{\delta \tau}{2} \sqrt{a^2-b^2}]$.
Similar expressions can be seen for other elements $M_2$. Thus up to
an overall irrelevant factor, our analysis reproduces the same
Mobius at every Trotter step as in Ref.\ \onlinecite{cft1}. A
similar analysis can be easily carried out for the heating phase and
leads to similar results.

Next, we show derive the form of $H_F$ using algebra of the Virasoro
operators without resorting to their SU(1,1) representations. To
this end, we begin from the time-dependent Hamiltonian given by Eq.\
\ref{hamcft}. For $f_0 \gg \delta f, 1$, the zeroth order evolution
operator is given by
\begin{eqnarray}
U_0(t,0) &=& \exp \left[- i L_0 \frac{2 \pi t}{L} \left(
\frac{f_0}{\omega_D t} \sin(\omega_D t) + \delta f\right) \right]
\label{ordzero}
\end{eqnarray}
This leads to $U_0(T,0)= \exp[-2 \pi i s L_0/L]$ and $H_F(0)= 2\pi s
L_0/(LT)$; these results coincide with Eq.\ \ref{zerothorder} for
$L_0 = \sigma_z/2$.

To obtain the next order correction to $U$, we write
\begin{eqnarray}
U'_1(T,0) &=& - i \int_0^T U_0^{\dagger}(t,0) \frac{\pi}{L}
(L_1+L_{-1}) U_0(t,0) dt  \nonumber\\ \label{fordcorr}
\end{eqnarray}
To evaluate this, we use the Baker-Campbell-Hausdorff relations for
$L_0$ and $L_{\pm 1}$ which states
\begin{eqnarray}
e^{i a L_0} (L_1+ L_{-1}) e^{-i a L_0} &=& e^{i a} L_1 + e^{-i a}
L_{-1} \label{bch1}
\end{eqnarray}
Using Eq.\ \ref{bch1} and identifying $a = - 2 \pi/L ( f_0
\sin(\omega_D t)/\omega_D + \delta f t)$ (Eq.\ \ref{ordzero}), one
can evaluate $U'_1 (T,0)$ in a straightforward manner. The integrals
involved are similar to those in App.\ \ref{appa} and one obtains
\begin{eqnarray}
U'_1(T,0) &=& i \alpha \left[ i (L_1+ L_{-1}) \sin s  \right. \nonumber\\
&& \left. + 2 \sin^2 s/2 (L_1-L_{-1}) \right] \label{u1primeeq}
\end{eqnarray}
where $\alpha$ is given by Eq.\ \ref{alphaeq} of the main text. Thus
the expression of the evolution operator, up to first order in
perturbation theory, is given by
\begin{eqnarray}
U_1(T,0) &=& U_0(T,0) (I- U'_1(T,0)) \label{u1firstorder} \\
&=& e^{-i (2\pi s/L) T L_0} [ I -i \alpha\{ 2 \sin^2 (s/2)
(L_1-L_{-1}) \nonumber\\
&& + i \sin s (L_1+L_{-1}) \} ] \nonumber
\end{eqnarray}
We note that this expression matches with Eq.\ \ref{u1eq} in the
main text for $L_0 = \sigma_z/2$ and $L_{\pm 1}= \mp \sigma{\pm}$.

Next we briefly comment on the unitarization procedure to be
followed here. Since $U_1$ is non-unitary within first order FPT, we
would like to unitarize it. This seems difficult without using the
SU(1,1) representation of the Virasoro operators. Here we therefore
concentrate on the high frequency regime, where the two routes to
unitarization of $U_1$, discussed in the main text, leads to
identical result. In the high frequency regime, $ s \to 0$ and $
\sin s/ s \to 1$. Furthermore, in this regime once can write $U_1
\simeq 1 - i H_F T/\hbar$. Thus expanding Eq.\ \ref{u1firstorder} in
powers of $s$ and retaining only first order terms, we find
\begin{eqnarray}
H_F \simeq \frac{\theta}{T}  [2 n_z L_0 + n_y (L_1+ L_{-1})/i]
\label{flL}
\end{eqnarray}
where $n_z$, $n_y$, and $\theta$ are given by Eq.\ \ref{fpt1} of the
main text. We note that $H_F$ coincides with that obtained within
first order FPT obtained in the main text with the identification
$L_0 = \sigma_z/2$ and $L_1+ L_{-1} = i \sigma_y$ and is also
consistent with the dynamic emergent symmetry discussed in the main
text.

\end{document}